\documentstyle[12pt]{article}
\newcommand{\Section}[1]%
{\section{#1}\setcounter{equation}{0}%
\setcounter{theorem}{0}}
\newtheorem{theorem}{Theorem}
\newtheorem{lemma}[theorem]{Lemma}
\newtheorem{coro}[theorem]{Corollary}

\newenvironment{proof}[1]%
{\par\noindent{\em #1:\ }}%
{~\rule{2mm}{2mm}\par\bigskip}
\newcommand{\Hilb}{{\cal H}}
\newcommand{\setC}{{\cal C}}
\newcommand{\setD}{{\cal D}}
\newcommand{\ret}{\nonumber \\}
\newcommand{\average}[1]{\langle #1 \rangle_\mu}
\newcommand{\microage}[1]{\langle #1 \rangle}
\newcommand{\abs}[1]{\left| #1 \right|}
\newcommand{\norm}[1]{\left\| #1 \right\|}

\font\titlefntbig=cmbx12 scaled \magstep4
\font\titlefnt=cmbx12 scaled \magstep2
\font\namefnt=cmr12 scaled \magstep1
\begin{document}
\newpage\thispagestyle{empty}
{\topskip 3cm
\begin{center}
{\titlefntbig
Ferromagnetism\\\vskip5mm
in the Hubbard Model\\}\vskip1cm
{\titlefnt
-- Examples from Models with Degenerate\\\vskip3mm
Single-Electron Ground States\\}
\vfil
{\namefnt Andreas Mielke}\\
\bigskip
{\em Institut f\"ur theoretische Physik,
Universit\"at Heidelberg,\\ Philosophenweg 19,
D-6900 Heidelberg, Federal Republic of Germany\\}
\bigskip\bigskip
{\namefnt Hal Tasaki}\\
\bigskip
{\em Department of Physics, Gakushuin University\\
Mejiro, Toshima-ku, Tokyo 171, Japan\\}
\end{center}
\par\vfil\vfil\noindent
Whether spin-independent Coulomb interaction can be the origin of
a realistic ferromagnetism in an itinerant electron system has
been an open problem for a long time.
Here we study a class of Hubbard models on decorated lattices,
which have a special property that the corresponding
single-electron Schr\"{o}dinger
equation has $N_{\rm d}$-fold degenerate ground states.
The degeneracy $N_{\rm d}$ is proportional to the total number of
sites $\abs{\Lambda}$.
We prove that the ground states of the models exhibit
feand sufficiently close to $\rho_0=N_{\rm d}/(2\abs{\Lambda})$, and
paramagnetism when the filling factor is
sufficiently small.
An important feature of the present work is that it provides
examples of three dimensional itinerant electron systems which are
proved to exhibit ferromagnetism in a finite range of the electron
filling factor.
(To appear in Commun. Math. Phys.)
\vfil}\newpage
%%%%%%%%%%%%%%%%%%%%%%%%%%%%%%%%%%%%%%%%
%%%%%%%%%%%%%%%%%%%%%%%%%%%%%%%%%%%%%%%%
\tableofcontents\newpage
%%%%%%%%%%%%%%%%%%%%%%%%%%%%%%%%%%%%%%%%
%%%%%%%%%%%%%%%%%%%%%%%%%%%%%%%%%%%%%%%%
\Section{Introduction}
\subsection{Ferromagnetism in the Hubbard model}
In some solids, electronic spins spontaneously align with each other
to form strong ferromagnetic ordering.
A familiar example is Fe, which maintains long range magnetic order
up to the Curie temperature, 1043K.
Given the fact that interactions between electrons in a solid are
almost spin-independent, the existence of such a strong order may
sound as a mystery.
As we shall describe below, this has indeed been an interesting open
problem in theoretical physics for quite a long time.

In 1928, Heisenberg \cite{Heisenberg} pointed out that the
spin-independent
Coulomb interaction between electrons, when combined with the
Pauli exclusion principle, can generate effective interaction
between electron spins.
Heisenberg's picture of ferromagnetism was that the relevant
electrons are mostly localized at atomic sites,
and their spin degrees of freedom interact with each other via
``exchange interaction''.
It has been realized, however, that his exchange interaction usually
has the sign which leads to antiferromagnetic interaction rather
than ferromagnetic one.
(See \cite{Herring1} for a review.)
Nevertheless, Heisenberg's idea still plays a fundamental role modern theories
of ferromagnetism.

A somewhat different approach to ferromagnetism, which was
originated by Bloch \cite{Bloch}, is to look for a mechanism of
ferromagnetism in which the itinerant nature of electrons, as well
as the Coulomb interaction and Pauli principle, play fundamental
roles.
This project, combined with sophisticated band-theoretic
techniques, has led to many approximate theories \cite{Herring2}.
A common feature of all these theories is that they are based on the
Hartree-Fock approximation ({\em i.e.\/}, mean field theory) and its
perturbative corrections.
Although such approximations can lead to reasonable conclusions in
some situations, they have serious disadvantage from a theoretical
point of view.
The basic strategy of the approximations is to treat electrons with
up and down spins as different species of particles, and then
introduce some self-consistency conditions.
By doing this, one severely destroys the original $SU(2)$ invariance
of the model and gets
 $Z_2$ invariant self-consistent equations.
The existence of
 ferromagnetism then reduces to a problem of
spontaneous breakdown of the discrete $Z_2$ symmetry, which is
essentially different from the original
problem, a spontaneous breakdown of the continuous $SU(2)$
symmetry.
As a consequence, the approximate theories give only two
ferromagnetic states with the net magnetization pointing up or
down, instead of expected infinitely many states with an arbitrary
direction of magnetization.
Since a continuous symmetry breaking is a very subtle phenomenon
in general,
results based on the Hartree-Fock approximation and its
improvements are not conclusive enough to
answer the fundamental question whether
spin-independent Coulomb interaction alone can be the origin of a
realistic ferromagnetism in an itinerant electron system.
We stress that such a critical point of view hby many physicists.
See, for example, the review of Herring \cite{Herring2}.

Given the subtlety of the problem, it is desirable to have
idealized models in which one can develop concrete scenarios for the
itinerant electron ferromagnetism.
The so called Hubbard model \cite{Hubbard,Kanamori} is a simple but
nontrivial model
suitable for developing such scenarios.
There has been a considerable amount of heuristic works
(mostly based on
the Hartree-Fock approximation and its improvements) devoted to
theories of ferromagnetism in the Hubbard model.
Since the literature is too large to catalogue here, we only refer to
the pioneering work of Kanamori \cite{Kanamori} and a review by
Herring \cite{Herring2}.
The Hartree-Fock approximation applied to the Hubbard model
yields the so called Stoner criterion for ferromagnetism
\cite{Herring2,Hubbard,Kanamori}.
It says that ferromagnetism occurs if $D_{\rm F}U>1$,
where $D_{\rm F}$ is the single particle-density of states at the
fermi level
and $U$ is the strength of the Coulomb repulsion.
Although this is certainly not true in general, large values of
$D_{\rm F}$ and/or $U$ determine the region in parameter space
where one may find ferromagnetism.

In 1965, the first rigorous example of ferromagnetism in the
Hubbard model was given by Nagaoka \cite{Nagaoka}, and
independently by Thouless
\cite{Thouless,Lieb1}.
It was proved that certain Hubbard models have ground states with
saturated magnetization when there is exactly one hole and the
Coulomb repulsion is infinite.
Recently, it was pointed out that the theorem extends to a general
class of models which satisfy a certain connectivity condition
\cite{Tasaki1}.
Whether the Nagaoka-Thouless ferromagnetism survives in the
models with finite density of holes and/or finite Coulomb repulsion
is a very interesting but totally unsolved problem
\cite{Douc
In 1989, Lieb proved an important theorem on general properties of
the ground states of the Hubbard model \cite{Lieb2}.
As a consequence of the theorem, he showed that a class of Hubbard
models on asymmetric bipartite lattices with finite Coulomb
interaction have ferromagnetic ground states at half filling.
It is sometimes argued that Lieb's examples represent
ferrimagnetism ({\em i.e.\/}, antiferromagnetism on a bipartite
lattice in
which the number of the sites in two sublattices are different)
rather than ferromagnetism.
Although this might be true when the Coulomb repulsion is
sufficiently large and the dimension is high enough, we believe
mechanisms underlying his examples are much richer in general
situations.

A more recent example is due to Mielke, who studied Hubbard models
on general line graphs \cite{Mielke0,Mielke1}.
A special feature of his models is that the single-electron
Schr\"{o}dinger equation corresponding to the Hubbard model has
highly
degenerate ground states.
He proved that these models with nonvanishing Coulomb interaction
have ground states with saturated magnetization when the electron
number is exactly equal to the dimension of degeneracy of the
single-electron ground states.
Mielke also extended his results to a finite range of electron filling
factor in certain two dimensional models \cite{Mielke2}.
See \cite{Mielke3} for the recent results for a more general class of
models.

The latest example is due to Tasaki \cite{Tasaki2}, who proved the
existence of ferromagnetism in a class of Hubbard models
with nonvanishing Coulomb interaction on
decorated lattices.
As in the Mielke's models, the class of models have highly
degenerate
ground states in the corresponding single-electron Schr\"{o}dinger
equation.
It was proved that the ground states of the Hubbard model exhibit
ferromagnetism in a ffactor.
The proof made use of a basis of the single-electron ground states
which satisfy certain locality and (reflection) positivity conditions,
and a percolation representation for physical quantities.
This work covered models in three dimensions as well.

A common feature in the examples of Lieb, Mielke, and Tasaki is that
each model has a completely degenerate band in the corresponding
single-electron spectrum, and the ferromagnetism (in interacting
many electron problem) is proved when the degenerate band is
exactly or nearly half-filled.
This situation, where $D_{\rm F}$ is infinite, is, in view of the
Stoner criterion, in some sense dual to the forementioned theorem
of Nagaoka where $U$ is infinite.
Possibility of ferromagnetism in systems with a degenerate
single-electron band has been discussed for a long
time since the pioneering work of Slater, Statz and Koster
\cite{Slater} in 1953.
But general consensus on whether ferromagnetism appears or not
has been lacking.
See Section VIII.2 of \cite{Herring2}.

%%%%%%%%%%%%%%%%%%%%%%%%%%%%%%%%%%%%%%%%
\subsection{Purpose of the present paper}
The purpose of the present paper is to give further discussions on
the ferromagnetism studied in \cite{Tasaki2}.
We do not only present all the detailed analysis announced (but
omitted) in \cite{Tasaki2}, but also offer a new proof of the main
theorem.
Since the new proof is considerably simpler and stronger than the
original one, we are able to remove some conditions required in the
previous paper.
We no longer need the (reflection) positivity condition or the
symmetry condition stated in \cite{Tasaki2}, and are able to treat
models on an arbitrary decorated lattice.

Our results and their proof demonstrate that there is a
mechanism generated by the Coulomb interaction which selects
ferromagnetic states as ground states.
The Hamiltonian''
(\ref{3.18}), where the Hamiltonian for the Coulomb interaction,
projected
onto a particular subspace, is reduced to that of the ferromagnetic
Heisenberg model.
This can be regarded as a rigorous example of a ``(super) exchange
interaction''
which is ferromagnetic.
The selection mechanism works most effectively when the
degenerate single-electron band is nearly half filled (in the
sense that the electron number is nearly equal to the dimension of
degeneracy in the single-electron ground states), but becomes
ineffective when the electron density is too small.

Although the models treated in the present paper are still artificial,
we hope that such a selection mechanism generally takes place in
a Hubbard model with a large density of states at the bottom of the
single-electron energy band.
Such a Hubbard model should exhibit ferromagnetism
for suitable electron filling factors when
the Coulomb
interaction is sufficiently large.
In other words, we hope the present examples to provide the
simplest models in a ``universality class'' of Hubbard
models which exhibit realistic and robust ferromagnetism.
The recent numerical and heuristic works of Kusakabe and Aoki
\cite{Kusakabe2} for closely related models (which were proposed
by the present authors) indeed suggest that the
ferromagnetism in the present models is robust and not pathological.
It is a challenging problem to extend our results to more general
situations.

It is encouraging that ferromagnetism observed in transient metals,
like Ni, might have similar properties as our ferromagnetism.
The 3d band of Ni has large (single-electron) density of states at the
top of the band, and the filling factor of the band is close to one
\cite{Kanamori}.
After the electron-hole transformation, the situation becomes quite
similar tWhen the number of electrons is equal to the degeneracy of the
single-electron ground states, our model might also resemble
certain ferromagnetic ionic crystals.

A reader might feel that the Hubbard models treated in the present
work have ``nonstandard'' hopping matrix elements (including the
next nearest neighbor hopping) when compared with the ``standard''
models with uniform nearest neighbor hoppings.
(See eq.(\ref{2.3}) and Fig.~1.)  It should be remarked
that, unlike in the lattice field theories, for example, the hopping
matrix in the Hubbard model need not be the
naive discretization of the Laplacian.
It is determined from overlap integrals between electron orbits, and
can be quite complicated in general.
For example, the typical three-band Hubbard model used
to study the normal state of the high-$T_{\rm c}$ superconductors
(see {\em e.g.} \cite{hanke} and the references therein)
contains next-nearest neighbor hoppings and on-site potentials
as well and is very similar to our two-dimensional model
described in Section 2.1.
Another important reason for studying a ``nonstandard'' model is the
remarkable ``nonuniversality'' of the Hubbard model.
These days it is suspected that the simplest Hubbard model with
uniform hopping does not exhibit neither ferromagnetism nor
superconductivity for reasonable parameter values.
This is the case in recent numerical and analytical results for the
Hubbard model in infinite dimensions \cite{janis}.
However the chance is big that a particular version of the Hubbard
model with specific hopping matrix ({\em i.e.\/}, band structure) and
filling
factor shows such interesting properties.
This is reminiscent of the rich nonuniversal behavior one observes
in the actual itinerant electron systems in nature.

The organization of the present paper is as follows.
In Secdiscuss simple construction of ferromagnetic ground states in
certain models.
In Section 2, we describe our results for a particular Hubbard model,
and discuss their physical consequences.
The physics of the present paper can be read off from these two
sections.
Section 3 is devoted to a proof of our main theorem which provides a
complete characterization of ground states.
In Section 4, we derive percolation representation for various
physical quantities, and prove rigorous upper and lower bounds,
which have direct physical significance.
In Section 5, we calculate correlation functions.
In Section 6, we derive a ``spin Hamiltonian'' and discuss about
spin wave excitations.

%%%%%%%%%%%%%%%%%%%%%%%%%%%%%%%%%%%%%%%%
\subsection{Hubbard models with degenerate single-electron ground
states}
In the present subsection, we will give preliminary discussions on a
special class of Hubbard models, in which one can easily construct
exact ground states which are ferromagnetic.
We shall also give some general definitions.

We take a finite lattice $\Lambda$ with $\abs{\Lambda}$ sites and
consider a Hubbard model on $\Lambda$.
Throughout the present paper, we denote by $\abs{S}$ the number of
elements in a set $S$.
The Hamiltonian is
\begin{equation}
H={H}_{\rm hop}+{H}_{\rm int},
\label{1.1}
\end{equation}
where
\begin{equation}
{H}_{\rm hop}= \sum_{ x,y\in
\Lambda  ,\sigma  =\uparrow  ,\downarrow } {
t}_{ xy} {c}_{x\sigma }^{\dagger }{c}_{y\sigma },
\label{1.2}
\end{equation}
and
\begin{equation}
{H}_{\rm int}= \sum_{ x\in
\Lambda } { U}_{ x} {n}_{x\uparrow
}{n}_{x\downarrow }.
\label{1.3}
\end{equation}
As usual ${c}_{x\sigma }^{\dagger }$ and ${c}_{x\sigma }$ are the
creation and the annihilation operators, respectively, of an electron
at site $x \in \Lambda $ with spin $\sigma
=\uparrow,\downarrThey satisfy the anticommutation relations
\begin{equation}
\{{c}_{x\sigma },{c}_{y\tau }^{\dagger }\}={\delta }_{xy}{\delta
}_{\sigma \tau },
\label{1.4a}
\end{equation}
 and
\begin{equation}
\{{c}_{x\sigma },{c}_{y\tau }\}=\{{c}_{x\sigma }^{\dagger
},{c}_{y\tau
}^{\dagger }\}=0,
\label{1.4b}
\end{equation}
for any $x,y\in\Lambda$ and $\sigma,\tau=\uparrow,\downarrow$,
where $\{A,B\}=AB+BA$.
The number operator is defined as
\begin{equation}
n_{x\sigma }={c}_{x\sigma }^{\dagger }{c}_{x\sigma }.
\label{1.5}
\end{equation}
The hopping matrix $(t_{xy})$ is real symmetric, and the on-site
Coulomb repulsion $U_x$ is strictly positive.
The total electron number operator is
\begin{equation}
\hat{N}_{\rm e}= \sum_{x\in\Lambda }
 (n_{x\uparrow }+n_{x\downarrow}),
\label{1.6}
\end{equation}
and we denote by $N_{\rm e}$ its eigenvalue.
A standard prescription is to consider the eigenspace of
$\hat{N}_{\rm e}$ with a given eigenvalue $N_{\rm e}$, or to
consider certain grand canonical ensemble with the expectation
value of $\hat{N}_{\rm e}$ fixed.
The quantity $N_{\rm e}/(2\abs{\Lambda})$ is called the electron
filling factor.

We also define the spin operators by
\begin{equation}
S_x^{(\alpha)} = \sum_{\sigma,\tau=\uparrow,\downarrow}
c^\dagger_{x\sigma} p^{(\alpha)}_{\sigma\tau} c_{x\tau}/2,
\label{1.7}
\end{equation}
where $p^{(\alpha)}_{\sigma\tau}$ with $\alpha=1,2,3$ are the Pauli
matrices,
\begin{equation}
p^{(1)}=\left(\begin{array}{cc}0&1\\1&0\end{array}\right),
\quad
p^{(2)}=\left(\begin{array}{cc}0&-i\\i&0\end{array}\right),
\quad
p^{(3)}=\left(\begin{array}{cc}1&0\\0&-1\end{array}\right).
\label{Pauli}
\end{equation}
We denote by $S_{\rm tot}(S_{\rm tot}+1)$ the eigenvalue of
\begin{equation}
({\bf S}_{\rm tot})^2 = \sum_{ x,y\in \Lambda }
\sum_{\alpha=1,2,3} S_x^{(\alpha)} S_y^{(\alpha)},
\label{1.8}
\end{equation}
which is the square of thWe say that a state exhibits ferromagnetism if it has
$S_{\rm tot}$
proportional to the system size $\abs{\Lambda}$.

The single-electron Schr\"{o}dinger equation corresponding to the
hopping Hamiltonian (\ref{1.2}) is
\begin{equation}
 \sum_{y\in\Lambda} { t}_{ xy}
{\varphi }_{y}=\varepsilon  {\varphi }_{x},
\label{1.9}
\end{equation}
where $\varphi_x\in C$, and $\varepsilon$ is the single-electron
energy.
Suppose that our hopping matrix $(t_{xy})$ has a special property
that the (single-electron) ground states of the Schr\"{o}dinger
equation
(\ref{1.9}) are $N_{\rm d}$-fold degenerate.
We denote the ground state energy by $\varepsilon_0$, and the space
of
degenerate ground states by $\Hilb_0$.
Let $\{\varphi^{(u)}\}_{u\in V}$ be a complete linear independent
basis for the
space $\Hilb_0$ ,
where $V$ (with $\abs{V}=N_{\rm d}$) is the set of indices.
The wave function and the creation operator corresponding to a basis
state ${\varphi }^{(u)}$ are denoted as
$\{\varphi_x^{(u)}\}_{x\in\Lambda}$ and
\begin{equation}
{a}_{u\sigma }^{\dagger }= \sum_{ x\in
\Lambda } { \varphi }_{ x}^{ (u)}
{c}_{x\sigma }^{\dagger },
\label{1.10}
\end{equation}
respectively.

Consider a ferromagnetic state
\begin{equation}
{ \Phi }_{A\uparrow } = \prod_{ u\in
A} { a}_{ u\uparrow }^{ \dagger }
{ \Phi }_0,
\label{1.11}
\end{equation}
 where $A$ is an arbitrary subset of the index set $V$,
and $\Phi_0$ is the vacuum state, {\em i.e.\/}, the state with no
electrons.
Since the basis $\{{\varphi }^{(u)}{\}}_{u\in  V}$ is linear
independent, the state ${ \Phi }_{A\uparrow }$ is nonvanishing.
The electron number of the state  ${ \Phi }_{A\uparrow }$ is given by
$N_{\rm e}=\abs{A}$.
By using the commutation relation
\begin{equation}
{H}_{\rm hop}{a}_{u\sigma }^{\dagger }={a}_{u\sigma }^{\dagger
}{H}_{\rm
hop}+{\varepsilon }_0 a^{\dagger}_{u\sigma},
\label{1\end{equation}
we find
$H_{\rm hop}\Phi_{A\uparrow }
= N_{\rm e}\varepsilon_0 \Phi_{A\uparrow }$,
where $N_{\rm e}\varepsilon_0$ is the lowest possible eigenvalue
of $H_{\rm hop}$ in the subspace with the electron number is fixed
to $N_{\rm e}$.
On the other hand, we already know from the construction that
$H_{\rm int} \Phi_{A\uparrow} = 0$, where $0$ is the minimum
possible eigenvalue of $H_{\rm int}$.
Therefore we have found the following.
\begin{theorem}
Consider a Hubbard model with the Hamiltonian described by
(\ref{1.1}),
(\ref{1.2}) and (\ref{1.3}).
In the subspace with the electron number fixed to ${N}_{\rm e}\le
N_{\rm d}$, the ground state energy is
${N}_{\rm e}{\varepsilon }_0$, and the ferromagnetic state
(\ref{1.11}) with an arbitrary subset $A\subset V$ with
$\abs{A}=N_{\rm e}$ is a ground state.
\label{trivialTheorem}
\end{theorem}

Such a construction of ferromagnetic ground states may be standard,
but is not sufficient to draw any meaningful conclusion about the
magnetism of the system.
A really important (and much more delicate) problem is whether
these ferromagnetic states are the only ground states, or what are
the other ground states, if any.
Note that the single-electron density of states $D_{\rm F}$ is
infinite for $N_{\rm e}\le N_{\rm d}$.
Therefore the Stoner criterion $D_{\rm F}U>1$ predicts the
appearance of only the ferromagnetic
ground states for any value of $U_x>0$.
But it soon turns out that the situation is not that simple.
A trivial counter example is the model with ${t}_{xy}=0$ for all
$x$, $y$, which has the degeneracy $N_{\rm d}=\abs{\Lambda}$.
Any state with no doubly occupied site is a ground state of this
model, so there can be no magnetic ordering.
We have paramagnetism, in contrast to the
conclusion based on the Stoner criterion.

In the present paper we shall study a class of Hubbard models, in
which
%%%%%%%%%%%%%%%%%%%%%%%%%%%%%%%%%%%%%%%%
%%%%%%%%%%%%%%%%%%%%%%%%%%%%%%%%%%%%%%%%
\Section{Main results and physical consequences}
\subsection{Definition}
In the present section, we discuss our rigorous results and their
physical consequences.
In order to simplify the discussion, we shall concentrate on the
simplest class of models defined on the decorated hypercubic
lattice.
Many of our results extend to other models with only minor
modifications.
See the remark at the end of the present subsection.

We shall begin with the definition of the model.
Consider a $d$-dimensional $L\times  \cdots \times  L$
hypercubic lattice, where $L$ is an even integer, and denote by $V$
the
set of sites.
We impose periodic boundary conditions.
(Note that, in section 1.2, we used the symbol $V$ to denote the
index set for the basis states.
The reason for using the same symbol for the set of sites will
become clear when we construct a basis at the end of Section~2.2.)
Let
\begin{equation}
B=\{\{v,w\}\left|\,v,w\in  V,\abs{v-w}=1\right.\},
\label{2.1}
\end{equation}
be the set of bonds, where $\abs{v-w}$ denotes the euclidean
distance between the sites $v$ and $w$.
For each bond $\{v,w\}$ in $B$, we denote by $m(v,w)$ the
point taken in the middle of the sites $v$ and $w$.
We define
\begin{equation}
M=\{m(v,w)\left|\,\{v,w\}\in B\right.\},
\label{2.2}
\end{equation}
and consider the decorated hypercubic lattice $\Lambda=V\cup M$.

We will study the Hubbard model on the lattice $\Lambda$.
We again write our Hamiltonian as $H={H}_{\rm hop}+{H}_{\rm int}$
where
\begin{equation}
{H}_{\rm hop}= \sum_{ \sigma
=\uparrow  ,\downarrow } \sum_{ \{v,w\}\in
  B}  t({c}_{v\sigma }^{\dagger }+{c}_{w\sigma
}^{\dagger }+\lambda  {c}_{m(v,w)\sigma }^{\dagger })({c}_{v\sigma
}+{c}_{w\sigma }+\lambda  {c}_{m(v,w)\sigma }),
\label{2.3\end{equation}
 and
\begin{equation}
{H}_{\rm int}=U \sum_{ u\in  V} { n}_{ u\uparrow }
{n}_{u\downarrow }+U'
\sum_{ x\in  M} { n}_{
x\uparrow } {n}_{x\downarrow },
\label{2.4}
\end{equation}
with $t>0$, $\lambda>0$, $U>0$, and $U'>0$.
The case  $\lambda<0$ is related to $\lambda>0$ by a unitary
transformation.
In the case $t<0$, the highest single-particle energy is degenerate.
This case can be mapped to $t>0$ by a particle-hole transformation.
Note that the above hopping Hamiltonian and interaction Hamiltonian
can be written in the form of (\ref{1.2}) and (\ref{1.3}) by suitably
choosing the
hopping matrix $({t}_{xy})$ and the interaction ${U}_{x}$.
See Fig.~1.

In general we can take an arbitrary finite lattice $V$,
and construct
the corresponding decorated lattice $\Lambda$ by adding points at
the center of each bond in $V$.
The Hamiltonian is defined as in (\ref{2.3}) and (\ref{2.4}), but the
parameters $t$, $\lambda$, $U$, and $U'$ can be bond (or site)
dependent, provided that they have the required signs.
Corollary~\ref{Theorem 2.2} and Theorem~\ref{half-filled theorem}
extend immediately to such general models.
If the dimension of $V$ is not less than two, and the coordination
number of $V$ is uniformly bounded, one can extend
Theorems~\ref{Theorem 2.3} and \ref{Theorem 2.4} with some extra
care.

%%%%%%%%%%%%%%%%%%%%%%%%%%%%%%%%%%%%%%%%
\subsection{Single-electron properties}
The special form of the hopping Hamiltonian (\ref{2.3}) makes the
present
model fall into the class considered in Section 1.2.
To see this, we should study the eigenstates of the single-electron
Schr\"{o}dinger equation (\ref{1.9}) corresponding to (\ref{2.3}).
A standard way is to use the Fourier transformation to directly
solve the eigenvalue problem.
One easily finds that the eigenstates can be classified into
$(d+1)$-bands, whose dispersion relations are given \begin{equation}
\varepsilon_i(k) = \left\{\begin{array}{ll}
0 & \mbox{for $i=1$},\\
\lambda^2 t & \mbox{for $i=2,3,\cdots,d$},\\
\lambda^2 t + 2t \sum_{j=1}^d(1+\cos k_j) & \mbox{for $i=d+1$},
\end{array}\right.
\label{band}
\end{equation}
where $i=1,2,\cdots,d+1$ is the index for the bands, and
$k=(k_1,\cdots,k_d)$ is the wave vector with $-\pi<k_j\le\pi$.
Note that the present model has a singular band structure, in which
most of the bands are dispersion-less.

Rather than making use of the direct Fourier transformation
calculation, however, we here make use of special features of
(\ref{2.3}) to get the following.
\begin{lemma}
Consider the single-electron Schr\"{o}dinger equation (\ref{1.9})
corresponding to the hopping Hamiltonian (\ref{2.3}).
Then the ground states have $\varepsilon = 0$, and are
characterized
by the condition that
\begin{equation}
{\varphi }_{v}+{\varphi }_{w}+\lambda\,{\varphi }_{m(v,w)}=0
\label{2.5}
\end{equation}
holds for all $\{v,w\}\in  B$.
The dimension $N_{\rm d}$ of the corresponding eigenspace
$\Hilb_0$ is equal to $\abs{V}=L^d$.
\label{Lemma 2.1}
\end{lemma}
\begin{proof}{Proof}
Since the operator ${H}_{\rm hop}$ is positive semidefinite, the
eigenvalues must satisfy $\varepsilon  \ge 0$.
Let us define a single-electron state as $ \Phi  =({
 \sum}_{x\in  \Lambda } {\varphi }_{x}{c}_{x\sigma }^{\dagger
}){ \Phi }_0$ and note that the Schr\"{o}dinger equation (\ref{1.9}) is
written as ${H}_{\rm hop} \Phi  =\varepsilon
\Phi $.
Noting that
\begin{eqnarray}
&&({c}_{v\sigma }^{\dagger }+{c}_{w\sigma }^{\dagger
}+\lambda  {c}_{m(v,w)\sigma }^{\dagger })({c}_{v\sigma
}+{c}_{w\sigma }+\lambda  {c}_{m(v,w)\sigma })(
\sum_{ x\in  \Lambda } { \varphi }_{
x} {c}_{x\sigma }^{\dagger }){ \Phi }_0
\ret
&&=({\varphi }_{v}+{\varphi }_{w}+\lambda  {\varphi
}_{m(v,w)})({c}_{v\sigma }^{\dagger }+{c}_{w\sigma }^{\dagger
}+\lambda
 \label{2.6}
\end{eqnarray}
we see that the condition (\ref{2.5}) gives a necessary and
sufficient condition for the state $\Phi$ to have  $\varepsilon=0$.
To get the dimension, one only has to note that the eigenspace is
determined by $\abs{\Lambda}-\abs{V}$ independent constraints.
\end{proof}

{}From Lemma~\ref{Lemma 2.1} and Theorem~\ref{trivialTheorem},
we get the following preliminary result about ferromagnetism.

\begin{coro}
In the subspace with the electron number fixed to ${N}_{\rm
e}\le N_{\rm d}(=L^d)$, the
ground state energy of the full Hubbard Hamiltonian $H$ (defined by
(\ref{2.3}) and (\ref{2.4})) is $0$.
Among the ground states, there are the ferromagnetic states defined
as (\ref{1.11}).
\label{Theorem 2.2}
\end{coro}

Let us construct a basis for the space $\Hilb_0$.
We shall label each basis state by a site $u$ in the hypercubic
lattice $V$.
For each $u\in V$, we define a single-electron state $\{{\varphi
}_{x}^{(u)}{\}}_{x\in  \Lambda }$ by
\begin{equation}
{\varphi }_{x}^{(u)}=\left\{{\begin{array}{ll}1&x=u\\
-{\lambda }^{-1}&x=m(u,v) \,\mbox{for some $v$}\\
0& \mbox{otherwise},\end{array}}\right.
\label{2.7}
\end{equation}
which clearly satisfies the condition (\ref{2.5}).
The basis is not orthogonal, but is easily checked to be linear
independent.
Certain characters of the basis will play central role in the proof of
the main theorem in
Section 3.

It should be stressed that the locality of our basis (\ref{2.7}) does
not
imply that electrons are localized in the present model.
One can always take a standard basis states with definite crystal
momenta, in which the single-particle states are extended.
The actual behavior of electrons should be determined by studying
various correlation functions.
See discussions in Section 2.3.

%%%%%%%%%%%%%%%%%%%%%%%%%%%%%%%%%%%%%%%%
\subsection{Fer\label{Ferro1}
Our first nontrivial result about ferromagnetism deals with the
model with a special electron number.

\begin{theorem}
In the subspace with the electron number fixed to $N_{\rm e} =
N_{\rm d} (= L^d)$, the ground states of $H$ have
$S_{\rm tot}=N_{\rm e}/2$ and are nondegenerate apart from the
$(2S_{\rm tot}+1)$-fold spin degeneracy.
\label{half-filled theorem}
\end{theorem}

The theorem will be proved in Section~3.1 as
Corollary~\ref{half filled}.
The theorem establishes that the ground states exhibit the maximum
possible ferromagnetism.
Mielke \cite{Mielke0,Mielke1} proved similar result for a general
class of Hubbard models on line graphs and on some decorated
graphs similar to our model but with additional hopping matrix
elements between the sites $m(v,w)$.
Recently Mielke \cite{Mielke3} extended his results to a general
class of Hubbard models with a degenerate single-electron ground
states.
The result of \cite{Mielke3} includes that of both
\cite{Mielke0,Mielke1} and the above Theorem~\ref{half-filled
theorem}.
See the remark at the end of Section~3.1 for further discussions.
Note that, in the band-theoretic language, the degenerate
single-electron ground state band (indexed as $i=1$ in (\ref{band}))
is
exactly half-filled when $N_{\rm e} = N_{\rm d}$.

Note that the above theorem applies to the model with $d=1$ as
well.
This does not contradict with the general result of Lieb and
Mattis \cite{Lieb3,Aizenman}, which inhibits ferromagnetic order in
one dimension,
since our model contains non-nearest-neighbor hoppings.

By substituting the definition (\ref{2.7}) of the basis state into the
expression (\ref{1.11}) of the ferromagnetic ground state, we find
that the ground state for $N_{\rm e} = N_{\rm d}$ is given by
\begin{equation}
\Phi_{V\uparrow} = \prod_{u\in V} (c^\dagger_{u\uparr- \lambda^{-1} \sum_{m;
\abs{u-m}=1/2} c^\dagger_{m\uparrow})
\Phi_0,
\label{half-filled GS}
\end{equation}
and its $SU(2)$ rotations.

When the parameter $\lambda$ is extremely large, the ground state
(\ref{half-filled GS}) has essentially one up electron at each site
of the hypercubic lattice $V$.
We will show in Section~5 that the coherence length in the ground
state is equal to $\lambda^{-1}$.
This is extremely short if
$\lambda\gg1$.
In this limit, our model resembles that of nearly localized electrons
(as in the Heisenberg's work \cite{Heisenberg}), and the origin of the
ferromagnetism may be interpreted as a ``super exchange
interaction'' via the nonmagnetic atom on the site between two
magnetic atomic sites.
We also expect that the model with $N_{\rm e}=N_{\rm d}$
describes a kind of Mott insulator,
at least when $\lambda\gg1$ and $U\gg1$.

When $\lambda\ll1$, on the other hand, the coherence length
$\lambda^{-1}$
becomes large, and it is no longer possible to regard the present
model as that of localized electrons.
It seems that even the simplest model with $N_{\rm e}=N_{\rm d}$
contains many interesting physics, which remain to be understood.

%%%%%%%%%%%%%%%%%%%%%%%%%%%%%%%%%%%%%%%%
\subsection{Ferromagnetism in a finite range of electron numbers}
Let us investigate whether the ferromagnetism established for the
special electron number is stable when the electron number is
changed.
We stress that any physically realistic model of
ferromagnetism should possess such stability.
All the theorems presented here will be proved in Section~4.

In Section~3.1, we will find that the ground states of $H$ are highly
degenerate for $N_{\rm e}<N_{\rm d}$.
To get a physically meaningful results, we have to consider the
average over the degenerate ground states.
Instead of fixing the electron number explicitly, we will employ the
grand canonical forvalue of the electron number by choosing appropriate
chemical
potential.
The reason for using the grand canonical formalism here is mainly
technical.

For an arbitrary operator $O$, we define the grand canonical-like
average by
\begin{equation}
\langle  O{\rangle }_{\mu }={\frac{{\rm Tr}[O \exp
(\mu \hat{N}_{\rm e}){P}_0]}{{\rm Tr} [\exp (\mu
\hat{N}_{\rm e}){P}_0]}},
\label{2.8}
\end{equation}
where ${P}_0$ is the orthogonal projection operator onto the
eigenspace of $H$ with the eigenvalue $0$.
It is expected that, by choosing a suitable (dimensionless) chemical
potential $\mu$ in (\ref{2.8}), we recover zero-temperature
properties of
the system with a desired electron filling factor.
If the electron filling factor had a pathological behavior as a
function of $\mu$, the use of the grand canonical formalism could
not be justified.
The following theorem guarantees that this is not the case.

\begin{theorem}
For arbitrary values of $\mu$, we have the upper and lower bounds
\begin{equation}
(1+\frac{e^{-\mu}}{2})^{-1}\ge \frac{\average{\hat{N}_{\rm
e}}}{N_{\rm d}}
\ge 3^{-d}(1+\frac{e^{-\mu}}{2})^{-1}.
\label{2.9}
\end{equation}
In the dimensions $d\ge2$, there are positive finite constants ${\mu
}_{1}$, ${c}_{1}$ which depend only on the dimension $d$,  and for
any $\mu\ge {\mu }_{1}$, we have the lower bound
\begin{equation}
{\frac{\langle  \hat{N}_{\rm e}{\rangle }_{\mu }}{N_{\rm d}}}\ge 1-
{c}_{1}{e}^{-
\mu }.
\label{2.10}
\end{equation}
\label{Theorem 2.3}
\end{theorem}

The above bounds determine the behavior of the electron filling
factor for extreme values of the chemical potential.
When ${\mu }$ is negative and its absolute value is large (compared
to $1$), (\ref{2.9}) implies
\begin{equation}
{\frac{\langle  \hat{N}_{\rm e}{\rangle }_{\mu }}{2|\Lambda
|}}\approx
 {e}^{\mu }.
\label{2.11}
\end{equation}
Wh\begin{equation}
{\rho }_0-{\frac{\langle  \hat{N}_{\rm e}{\rangle }_{\mu
}}{2|\Lambda
|}}\approx  {e}^{-\mu },
\label{2.12}
\end{equation}
where the maximum value of the electron filling factor is defined as
$\rho_0=N_{\rm d}/(2\abs{\Lambda})=(2d+2)^{-1}$.
The relation $\approx $ means that the both sides behave equally
apart from multiplication by a uniformly bounded function of
$\mu$.

Now we can state our main theorem.

\begin{theorem}
In the dimensions $d\ge2$, there are finite constants ${c}_{2}$,
${c}_{3}$, ${\mu }_{1}$, ${\mu }_{2}$ (with ${c}_{2}, c_3,\mu_1>0$
and ${\mu }_{2}<0$) which depend only on the dimension $d$ and not
on the size of
the lattice.
For any $\mu\ge {\mu }_{1}$, we have
\begin{equation}
{S}_{ {\rm max}}({S}_{ {\rm max}}+1)\ge \langle  ({\bf  S}_{ \rm
tot}{)}^{2}{\rangle }_{\mu }\ge {S}_{\rm max}({S}_{\rm max}+1)(1-
{c}_{2}{e}^{-\mu }),
\label{2.13}
\end{equation}
where ${S}_{ {\rm max}}=N_{\rm d}/2$.
For any $\mu\le{\mu }_{2}$, we have
\begin{equation}
{\frac{3}{4}}\langle  \hat{N}_{\rm e}{\rangle }_{\mu }\le \langle
({\bf  S}_{\rm tot}{)}^{2}{\rangle }_{\mu }\le
{\frac{3}{4}}\langle
 \hat{N}_{\rm e}{\rangle }_{\mu }+{c}_{3}|V|{e}^{2\mu }.
\label{2.14}
\end{equation}
\label{Theorem 2.4}
\end{theorem}

Note that, when the bounds (\ref{2.13}) hold, the total spin of the
model is
proportional to the number of sites $\abs{\Lambda}$.
When the bounds (\ref{2.14}) hold, on the other hand, the total spin
is
proportional to the square root of $\abs{\Lambda}$.
Therefore Theorem \ref{Theorem 2.4} establishes that the ground
states of our
Hubbard model exhibit ferromagnetism when the filling factor is not
more than and sufficiently close to ${\rho }_0$, and paramagnetism
when the filling factor is sufficiently small.
In the band-theoretic language, ferromagnetism apdegenerate ground state band
is nearly half-filled.

The requirement that the dimension is not less than two in
Theorem~\ref{Theorem 2.4} is essential in controlling
ferromagnetic region by using a kind
of Peierls argument.
(See Section 4.4.)
The paramagnetic part (the bounds (\ref{2.14})) of
Theorem~\ref{Theorem 2.4} easily extends to the case with $d=1$.
We believe that the one-dimensional model exhibits paramagnetism
for all the values of $\mu<\infty$.

As we have discussed at the end of Section~1.3,
the Stoner criterion from the Hartree-Fock approximation predicts
the appearance of only the ferromagnetic ground states for all
$\mu<\infty$.
Theorem~\ref{Theorem 2.4} clearly shows that this is not the case.
For selection of ferromagnetic states to take place, the degenerate
single-electron band must be nearly half-filled in the sense that
$0<\rho_0-\rho\ll1$.
We expect that this feature is universal in the Hubbard models with
large single-electron density-of-states.
It is interesting that, in perturbative corrections to the
Hartree-Fock approximation, one finds somewhat similar
conditions in order
to ensure sufficiently large ``effective'' $U$.
See \cite{Kanamori} and section X of \cite{Herring2}.
See also \cite{Kusakabe} for a related discussions.

It is an interesting problem whether the ferromagnetism in the
present rather artificial model is stable under various perturbations
to the Hamiltonian.
We believe that the ferromagnetism persists under small
perturbations if the on-site Coulomb repulsion is sufficiently large,
but we have no rigorous results at the moment.
The numerical results of \cite{Kusakabe2} for
closely related models (one of which is introduced in \cite{Mielke1} and
the other is defined at the end of
Section~3.1 of the present paper) indicate that the present
ferromagnetism is stable
under perturbations.
It in the three dimensional model is
present at finite temperatures.
Recall that, in one and two dimensions, ferromagnetic order in any
Hubbard model is destroyed by thermal fluctuation at finite
temperatures \cite{Ghosh,Koma}.

%%%%%%%%%%%%%%%%%%%%%%%%%%%%%%%%%%%%%%%%
\subsection{Other properties}
In Subsection~\ref{Ferro1}, we argued that the model with the
special electron filling factor $\rho=\rho_0$ (which corresponds to
$N_{\rm e}=N_{\rm d}$) is likely to be an insulator (at least when
$\lambda\gg1$ and $U\gg1$).
When the filling factor $\rho=\average{\hat{N}_{\rm
e}}/(2\abs{\Lambda})$ is strictly less than $\rho_0$, on the other
hand, band-theoretic intuition suggests that the model describes a
metal.

Let ${E}_{\rm GS}({N}_{\rm e})$ be the ground state energy of a
Hubbard model with the electron number fixed to ${N}_{\rm e}$.
The charge gap of the model is defined as
\begin{equation}
{\Delta }_{\rm charge}({N}_{\rm e})={E}_{\rm GS}({N}_{\rm e}+1)-
2{E}_{\rm GS}({N}_{\rm e})+{E}_{\rm GS}({N}_{\rm e}-1).
\label{2.15}
\end{equation}
It is believed that, if a system has nonvanishing charge gap (in the
infinite volume limit), it describes an insulator.
In our case, we already know from Theorem~\ref{Theorem 2.2} that
${E}_{\rm
GS}({N}_{\rm e})=0$ for any ${N}_{\rm e}\le N_{\rm d}$, which
means
that ${\Delta }_{\rm charge}({N}_{\rm e})=0$ for any ${N}_{\rm
e}\le N_{\rm d}-1$.
This might first appear as an indication that the present model
describes a ferromagnetic metal when the electron filling factor
is strictly smaller than ${\rho }_0$.
However one must recall that vanishing of the charge gap is a
necessary but not a sufficient condition for an electron system to be
a conductor.
(A trivial example of an insulator with a vanishing charge gap is the
model with ${H}_{\rm hop}=0$.)

To deteelaborate analysis on the transport properties seems necessary.
At present we are not able to calculate conductivity (from, e.g., the
Kubo formula).
We can only calculate certain static correlation functions in the
model with $N_{\rm d}-2d+1 \le N_{\rm e} \le N_{\rm d}$.
See Section~5.

A rotation invariant system with spontaneous ferromagnetic order
always has spin wave excitations.
In Section~6, we show that the interaction Hamiltonian (\ref{2.4}),
projected onto a certain subspace, becomes
\begin{equation}
P_A H_{\rm int} P_A = 2 J(\lambda) \sum_{\langle u,v\rangle}
\left(\frac{1}{2}-2\tilde{\bf S}_u\cdot\tilde{\bf S}_v\right)P_A,
\label{spinHam}
\end{equation}
where the summation is over nearest neighbor pairs of sites in $V$,
and $J(\lambda)>0$.
(See Section~6 for definitions.)
It is remarkable that (\ref{spinHam}) is nothing but the Hamiltonian
of the ferromagnetic Heisenberg model.
We expect that the low energy spin excitations of the model are
approximately described by the ``spin Hamiltonian'' (\ref{spinHam}).
Then there should be spin wave excitations when the ground states
exhibit long range order.
When $N_{\rm e}=N_{\rm d}$ ({\em i.e.}, the degenerate band is
exactly half-filled), we prove that
there are excited states whose excitation
energies are bounded from above by that of the spin wave
excitations of the Heisenberg model.
See \cite{Kusakabe2} for discussions on low-lying excitations
 in the closely
related Hubbard model with $U=\infty$.

%%%%%%%%%%%%%%%%%%%%%%%%%%%%%%%%%%%%%%%%
%%%%%%%%%%%%%%%%%%%%%%%%%%%%%%%%%%%%%%%%
\Section{Characterization of ground states}
\subsection{Main theorem}
In the present section, we shall state and prove our main theorem.
The theorem provides a complete characterization of the ground
states when the electron number $N_{\rm e}$ does not exceedegeneracy $N_{\rm
d}$ of the single-electron ground states.
This result will be used in Section 4 to prove various bounds for
physical quantities.
The class of models treated here, which includes the model of
Section 2 as a
special case, is specified by a set of conditions for a basis
$\{\varphi^{(u)}\}_{u\in V}$ of the space $\Hilb_0$ of degenerate
single-electron ground states.
See B0, B1, and B2 below.

Let $\Lambda$ be a finite lattice.
We consider a Hubbard model on $\Lambda$ with the Hamiltonian
defined by
(\ref{1.1}), (\ref{1.2}) and (\ref{1.3}).
As in Section 1.2, we denote by $\{\varphi^{(u)}\}_{u\in V}$
a basis for the $N_{\rm d}$-dimensional space $\Hilb_0$ of the
single-electron ground states,
where $V$ is the index set.
Recall that, in Section 2, the symbol $V$ stands both for the
(undecorated) hypercubic lattice and for the index set.
We stress that this is an accidental coincidence in the model of
section 2, and $V$ is not necessarily a set of sites in the present
section.

We denote by ${a}_{u\sigma }^{\dagger }$ the creation operator
defined as in (\ref{1.10}), which corresponds to a basis state
${\varphi }^{(u)}$.
We have seen in Section 1.2 that the ferromagnetic state
(\ref{1.11}) is an
exact ground state of the model.
We start from explicit construction of other ground states.

We introduce a notion of connectivity in the index set $V$ by
declaring that two indices $u$ and $v$ are directly connected if
$\varphi_x^{(u)}\varphi_x^{(v)}\ne0$ for some site $x\in\Lambda$.
Let $A$ be an arbitrary subset of the index set $V$.
The subset $A$ can be uniquely decomposed into a disjoint
union of connected components as $A = {C}_{1}\cup  \cdots \cup
 {C}_{n}$.
Note that, in the ground state (\ref{1.11}), electrons on different
connected components may be regarded as not interacting with each
other.
(Of course this is a basis-dependebe taken literally.)
For each $k=1,2,\cdots,n$, let us define a
subset $\Lambda_k$ of the lattice $\Lambda$ by
\begin{equation}
\Lambda_k=\left\{x\in\Lambda\left|\,\varphi_x^{(u)}\ne0
\quad\mbox{for some $u\in{C}_k$}\right.\right\}.
\label{3.4}
\end{equation}
By the definition of connectivity, it immediately follows that
${\Lambda }_k\cap  {\Lambda }_{k'}=\emptyset$ for $k\ne k'$.
We define the spin lowering operator on ${\Lambda }_k$ by
\begin{equation}
{\hat{S}}_k^{-}= \sum_{ x\in  {\Lambda
}_k} { S}_{ x}^{ -}.
\label{3.5}
\end{equation}
Because of disjointness of the support sets, operators
${\hat{S}}_k^{-}$ with different $k$ commute with each other.
For $\{m_k{\}}_{k=1,\cdots ,n}$ with $m_k=-
|{C}_k|/2,1-|{C}_k|/2,\cdots ,|{C}_k|/2$, we define
\begin{equation}
{ \Phi }_{A,\{m_k\}} =\{ \prod_{
k=1}^n  ({\hat{S}}_k^{-}{)}^{(|{C}_k|/2)-
m_k}\}{ \Phi }_{A\uparrow },
\label{3.6}
\end{equation}
where
$\Phi_{A\uparrow }=\prod_{u\in A}a^\dagger_{u\uparrow}\Phi_0$
is the ferromagnetic ground state
defined in (\ref{1.11}).
Note that $m_k$ can be regarded as the total ${S}^{z}$ on the
sublattice ${\Lambda }_k$.

\begin{lemma}
In the subspace with the electron number fixed to $N_{\rm e}\le
N_{\rm d}$, the state ${ \Phi }_{A,\{m_k\}}$ with arbitrary
$A\subset V$ (such that $\abs{A}=N_{\rm e}$) and with arbitrary
$\{m_k{\}}_{k=1,\cdots ,n}$ is a ground state of the Hubbard model.
The ground state energy is equal to ${N}_{\rm e}{\varepsilon }_0$
where ${\varepsilon }_0$ denotes the single-electron ground state
energy.
\label{Lemma 3.1}
\end{lemma}
\begin{proof}{Proof}
For $u\in C_k$, we have
\begin{equation}
{\hat{S}}_k^{-}{a}_{u\uparrow }^{\dagger }={a}_{u\uparrow }^{\dagger
}{\hat{S}}_k^{-}+{a}_{u\downarrow }^{\dagger },
\label{3.7}
\end{equation}
which means that the states (\ref{3.6}) are still a linear
combinations of thstates $({ \prod}_{u\in A} {a}_{u\sigma
 (u)}^{\dagger }){ \Phi }_0$ with
$\sigma  (u)=\uparrow,\downarrow $.
This proves that ${H}_{\rm hop}{ \Phi }_{A,\{m_k\}}
={N}_{\rm e} {\varepsilon }_0 \Phi_{A,\{m_k\}}$.
Note that $N_{\rm e}{\varepsilon }_0$ is the lowest possible
eigenvalue of ${H}_{\rm hop}$.
On the other hand the identity ${n}_{x\uparrow }{n}_{x\downarrow
}{S}_x^{-} = 0$ and the fact that ${n}_{x\uparrow
}{n}_{x\downarrow }{ \Phi }_{A\uparrow } = 0$
immediately imply ${n}_{x\uparrow }{n}_{x\downarrow }{ \Phi
}_{A,\{m_k\}} = 0$ for any $x\in\Lambda$, and hence ${H}_{\rm
int}{ \Phi }_{A,\{m_k\}} = 0$.
Again $0$ is the lowest possible eigenvalue of $H_{\rm int}$.
\end{proof}

The ground states ${ \Phi }_{A,\{m_k\}}$ are not in general
eigenstates of the square of the total spin operator
$({\bf  S}_{\rm tot})^2$.
One can construct eigenstates by taking suitable linear
combinations.
For example, suppose that $A$ consist of two connected components
as $A={C}_{1}\cup  {C}_{2}$, and $|{C}_{1}|=|{C}_{2}|= 2 M$.
Then the state
\begin{equation}
 \sum_{ m=-M}^{ M}  (-1)^m\Phi_{A,\{m,-m\}}
\label{3.8}
\end{equation}
is a spin-singlet, {\em i.e.\/}, an eigenstate of
$({\bf S}_{\rm tot})^2$ with the vanishing eigenvalue.
Note that such construction of spin singlet ground states is possible
when $\abs{V}-{N}_{\rm e}$ is at least of order $L^{d-1}$ (where $L$
is the linear size of the lattice).
See \cite{Mielke1,Mielke2} for related results.

In order to completely characterize the set of ground states, we
require the basis $\{\varphi^{(u)}\}_{u\in V}$ to satisfy the
following three conditions.

\bigskip\noindent
{\bf B0.}
$\{\varphi^{(u)}\}_{u\in V}$ is a linear independent complete basis
of the space $\Hilb_0$.
Thus we have $\abs{V}=N_{\rm d}$.
For each $u\in V$, the corresponding wave function
$\{\varphi_x^{(u)}\}_{x\in\Lambda}$ is real.

\bigskip\no{\bf B1.} {\em Quasi locality:\/}
For each $u\in V$, there is a site $x(u)\in\Lambda $ with the
properties that $\varphi_{x(u)}^{(u)}\ne0$ and
$\varphi_{x(u)}^{(v)}=0$  for any $v\ne u$.

\bigskip\noindent
{\bf B2.} {\em Local connectivity:\/}
For each $x\in\Lambda$, there are at most
two indices $u\in V$ such that $\varphi_x^{(u)}\ne 0$.

\bigskip
Then the main theorem of the present section,
which will be proved in Section~3.2, is the following.

\begin{theorem}
Consider a Hubbard model with the Hamiltonian described by
(\ref{1.1}),
(\ref{1.2}) and (\ref{1.3}).
Let ${\varepsilon }_0$ and $\Hilb_0$ be the ground state energy and
the space of the ground states, respectively, of the corresponding
single-electron Schr\"{o}dinger equation (\ref{1.9}).
Suppose that one can find a basis $\{{\varphi }^{(u)}{\}}_{u\in
V}$ for the space $\Hilb_0$ which satisfies the conditions B0, B1,
and B2 stated above.
In the subspace with the electron number fixed to ${N}_{\rm e}\le
N_{\rm d}$, the ground state energy is
${N}_{\rm e}{\varepsilon }_0$, and an arbitrary
ground state is a linear combination of the states (\ref{3.6}) with
various
$A$ (with $\abs{A}=N_{\rm e}$) and $\{m_k\}$.
In the subspace with the electron number fixed to ${N}_{\rm e}>
N_{\rm d}$, the ground state energy is strictly larger than ${N}_{\rm
e}{\varepsilon }_0$.
\label{Theorem 3.2}
\end{theorem}

Mielke \cite{Mielke2} also gave a somewhat similar complete
characterization
of the ground states for a class of Hubbard models on
two-dimensional line graphs.

Theorem~\ref{Theorem 3.2} alone is not enough to determine
magnetic properties of the system.
As we have seen by constructing the state (\ref{3.8}), there are
states with small total spins among the degenerate ground states.
Whether the system exhibits ferromagnetism depends on how large
connected components of a ``typical'' subset $AWe shall deal with this problem
in Section~4.
The following special case, which corresponds to the ``half-filled
degenerate band'', can be treated without further analysis.

\begin{coro}
Assume the conditions for Theorem~\ref{Theorem 3.2}, and that the
index set $V$ is connected.
In the subspace with the electron number fixed to $N_{\rm
e}=N_{\rm d}$, the ground states of the Hubbard model have $S_{\rm
tot}=N_{\rm e}/2$, and are nondegenerate apart form the $(2S_{\rm
tot}+1)$-fold spin degeneracy.
\label{half filled}
\end{coro}
\begin{proof}{Proof}
We have $A=V$ because $\abs{A}=\abs{V}$.
Since $A$ is connected, (\ref{3.6}) represents fully ferromagnetic
states.
\end{proof}

\noindent {\bf Remark:}
Corollary~\ref{half filled}, which establishes
the existence of ferromagnetism for a
special electron number, can be proved in more general settings.
In \cite{Mielke1}, the same statement was proved for the Hubbard
models on general line graphs.
Let us mention that it is always possible to find a basis that
satisfies the conditions B0 and B1. This fact implies that the
connectivity of $V$ is necessary and sufficient for the
ferromagnetic ground state to be unique in the case
$N_{\rm e}=N_{\rm d}$. A proof and a basis-independent formulation
of this result was given in \cite{Mielke3}.

We shall briefly describe a class of models similar to the present
ones, in which the existence of ferromagnetism can be proved (at least)
for special electron numbers.
Let the unit cell $C$ consist of $n$ {\em external\/} sites
$x_1,\ldots, x_n$, and one {\em internal\/} site $y$.
(The cell $C$ can be, e.g., a bond, a plaquette, or a cube.
Then the external sites are corners, and the internal site is taken at
the center of $C$.
When $C$ is a bond, the models reduce to the ones considered in the
present paper.
One of the models studied in \cite{Kusakabe2} is oletting $C$ a plaquette.)
We define the hopping Hamiltonian within the cell as
\begin{equation}
H_{\rm hop}[C] = t\sum_{\sigma=\uparrow,\downarrow}
\left(\sum_{i=1}^nc^\dagger_{x_i\sigma}
+\lambda c^\dagger_{y\sigma}\right)
\left(\sum_{i=1}^nc_{x_i\sigma}
+\lambda c_{y\sigma}\right),
\end{equation}
where $t,\lambda>0$.
The whole lattice $\Lambda$ is constructed by assembling together
identical copies of the unit cell $C$.
When doing this, an external site may be (or may not be) shared by
several distinct unit cells, but an internal site should belong to
exactly one unit cell.
The total hopping Hamiltonian is
\begin{equation}
H_{\rm hop}=\sum_j H_{\rm hop}[C_j],
\label{generalHop}
\end{equation}
where $j$ is the index for the copies of the unit cell.
It is easily seen that the single electron ground states have  energy
$\varepsilon_0=0$, and their degeneracy $N_{\rm d}$ is identical to
the number of external sites, where an external site which belongs
to several different cells is counted only once.
When the whole lattice is connected, it can be proved that the
Hubbard model (with $U_x>0$) with the hopping Hamiltonian
(\ref{generalHop}) has fully polarized ferromagnetic ground states
for $N_{\rm e}=N_{\rm d}$.
The proof can be done by generalizing the one in the present paper,
but it follows immediately from the theorem in \cite{Mielke3}.

One can also introduce more complicated cell structures and
consider various arrangements of cells.
The existence of ferromagnetism in the resulting Hubbard model can
be proved by the theorem of \cite{Mielke3}.

%%%%%%%%%%%%%%%%%%%%%%%%%%%%%%%%%%%%%%%%
\subsection{Proof}
Following \cite{Mielke2},
we first introduce some creation and annihilation operators.
The Gramm matrix $G$ for the basis $\{\varphi^{(u)}\}_{u\in V}$ is
defined as
\begin{equation}
(G)_{uv} = \sum_{x\in\Lambda} \varphi_x^{\label{Guv}
\end{equation}
Since $\{\varphi^{(u)}\}_{u\in V}$ is linearly independent, the matrix
$G$ is regular.
We define
\begin{equation}
\kappa_x^{(u)} = \sum_{v\in V} (G^{-1})_{uv} \varphi_x^{(v)}.
\label{kappa}
\end{equation}
Note that $\{\{\kappa_x^{(u)}\}_{x\in\Lambda}\}_{u\in V}$ also
forms a basis of the space $\Hilb_0$.
It is easily verified that there are completeness conditions
\begin{equation}
\sum_{x\in\Lambda} \kappa_x^{(u)} \varphi_x^{(v)} = \delta_{uv},
\label{completeness1}
\end{equation}
and
\begin{equation}
\sum_{u\in V} \kappa_x^{(u)} \varphi_y^{(u)}
= \delta_{xy} - \psi_{xy},
\label{completeness2}
\end{equation}
where
\begin{equation}
\psi_{xy}=\psi_{yx}=\delta_{xy}
-\sum_{u,v\in V} \varphi_x^{(u)}(G^{-1})_{uv}\varphi_y^{(v)}.
\label{psixy}
\end{equation}
Note that $\psi_{xy}$ with a fixed $x$ (or $y$) defines a state
orthogonal to the space $\Hilb_0$ since
\begin{equation}
\sum_{y\in\Lambda} \psi_{xy}\varphi_y^{(u)} =
\sum_{y\in\Lambda} \psi_{xy}\kappa_y^{(u)}= 0,
\label{psiphi=0}
\end{equation}
for any $x\in\Lambda$ and $u\in V$.

Let us define
\begin{equation}
b_{u\sigma}=\sum_{x\in\Lambda}\kappa_x^{(u)}c_{x\sigma}
\label{bdef}
\end{equation}
and
\begin{equation}
d_{x\sigma}=\sum_{y\in\Lambda}\psi_{xy}c_{y\sigma}.
\end{equation}
{}From (\ref{completeness1}), we find that $b_{u\sigma}$ is the dual
of $a^\dagger_{u\sigma}$ in the sense that
\begin{equation}
\left\{b_{u\sigma}, a^\dagger_{v\tau}\right\}
= \delta_{uv}\delta_{\sigma\tau}.
\label{ba are dual}
\end{equation}
for any $u,v\in V$ and $\sigma,\tau=\uparrow,\downarrow$.
The orthogonality (\ref{psiphi=0}) implies
\begin{equation}
\left\{d_{x\sigma}, a^\dagger_{u\tau}\right\}
= \left\{d^\dagger_{x\sigma}, b_{u\tau}\right\} = 0,
\label{d anticommutes}
\end{equation}
for any $x\in\Lambda$, $u\in V$ and
$\sigma,\tau=\uparrow,\downarrow$.
Finally from tget the expansion formulas
\begin{eqnarray}
c^\dagger_{x\sigma} &=& \sum_{u\in V}
\kappa_x^{(u)}a^\dagger_{u\sigma}+d^\dagger_{x\sigma},
\label{expcdag}\\
c_{x\sigma} &=& \sum_{u\in V}
\varphi_x^{(u)}b_{u\sigma}+d_{x\sigma},
\label{expc}
\end{eqnarray}
for any $x\in\Lambda$ and $\sigma,\tau=\uparrow,\downarrow$.
The formulas (\ref{expcdag}) and (\ref{expc}) will turn out to be
useful later.

We shall now prove Theorem~\ref{Theorem 3.2}.
We restrict ourselves to the subspace with the electron number
fixed to $N_{\rm e}\le N_{\rm d}$.
By construction, we already know that the states (\ref{3.6}) are
ground states of the Hubbard model, and the ground state energy is
$N_{\rm e}\varepsilon_0$.
We will show that they are the only ground states.

Let $\Phi$ be a ground state.
Since a ground state satisfies $H_{\rm hop}\Phi = N_{\rm
e}\varepsilon_0 \Phi$, it can be written in the form
\begin{equation}
\Phi = \sum_{A_\uparrow,A_\downarrow}
f(A_\uparrow,A_\downarrow)
\prod_{u\in A_\uparrow} a^\dagger_{u\uparrow}
\prod_{v\in A_\downarrow} a^\dagger_{v\downarrow}\Phi_0,
\label{basicRep}
\end{equation}
where $A_\uparrow$ and $A_\downarrow$ are subsets of $V$ such
that $\abs{A_\uparrow}+\abs{A_\downarrow}=N_{\rm e}$, and
$f(A_\uparrow,A_\downarrow)$ is a coefficient.

A ground state must also satisfy $H_{\rm int}\Phi=0$, and hence
$c_{x\uparrow}c_{x\downarrow}\Phi = 0$ for any $x\in\Lambda$.
By substituting the expansion formula (\ref{expc}), this necessary
condition can be rewritten as
\begin{equation}
\left(\sum_{u\in V} \varphi_x^{(u)}b_{u\uparrow}\right)
\left(\sum_{v\in V} \varphi_x^{(v)}b_{v\downarrow}\right)
\Phi = 0,
\label{basicCond}
\end{equation}
for any $x\in\Lambda$.
We have used $d_{x\sigma}\Phi=0$, which follows from the
anticommutation relation (\ref{d anticommutes}) and the
representation (\ref{b
We shall first set $x=x(u)$, where $x(u)$ is the site defined in
the {\em quasi locality\/} condition B1.
Then the condition (\ref{basicCond}) becomes
\begin{equation}
(\varphi_{x(u)}^{(u)})^2b_{u\uparrow}b_{u\downarrow}\Phi=0.
\label{x(u)}
\end{equation}
By using the anticommutation relations (\ref{ba are dual}), we find
from (\ref{x(u)}) that the coefficients in (\ref{basicRep}) must
satisfy
\begin{equation}
f(A_\uparrow,A_\downarrow) = 0
\label{f=0}
\end{equation}
if $A_\uparrow\cap A_\downarrow\ne\emptyset$.

Let $u,v\in V$ be directly connected, and take $x\in\Lambda$ such
that $\varphi_x^{(u)}\varphi_x^{(v)}\ne0$.
By using the {\em local connectivity\/} B2 and (\ref{f=0}), we see
that the condition (\ref{basicCond}) becomes
\begin{equation}
\varphi_x^{(u)}\varphi_x^{(v)}
\left( b_{u\uparrow} b_{v\downarrow} -
b_{u\downarrow} b_{v\uparrow} \right) \Phi = 0.
\end{equation}
Again from the anticommutation relations (\ref{ba are dual}), we
find that
\begin{equation}
f(B_{\uparrow}\cup\{u\},B_{\downarrow}\cup\{v\})
= f(B_{\uparrow}\cup\{v\},B_{\downarrow}\cup\{u\}),
\label{f=f}
\end{equation}
where $B_{\uparrow}$ and $B_{\downarrow}$ are arbitrary subsets
of $V\backslash\{u,v\}$ such that
$\abs{B_{\uparrow}}+\abs{B_{\downarrow}}=N_{\rm e}-2$.
In other words, the coefficients $f(A_\uparrow,A_\downarrow)$ are
symmetric under the exchange of spins corresponding to the indices
$u$ and $v$.

Take a subset $A\subset V$ with $\abs{A}=N_{\rm e}$, and
decompose it into connected components as $A=\bigcup_{k=1}^n
C_k$.
Take $\{m_k{\}}_{k=1,\cdots ,n}$ with $m_k=-
|{C}_k|/2,1-|{C}_k|/2,\cdots ,|{C}_k|/2$.
By noting that the condition (\ref{f=f}) is valid for any pair of
directly connected indices $u$, $v$, we find that the coefficient
$f(A_\uparrow,A_\downarrow)$ takes a constant value for those
$A_\uparrow$, $A_\downarrow$ such that $A_\A_\downarrow = \emptyset$,
$A_\uparrow \cup A_\downarrow =
A$, and $\abs{A_\uparrow\cap C_k}-\abs{A_\downarrow\cap
C_k}=2m_k$ for any $k=1,\cdots,n$.
This proves that any $\Phi$ of the form (\ref{basicRep}) satisfying
the conditions (\ref{f=0}) and (\ref{f=f}) is a linear combination of
the states (\ref{3.6}).
The first part of the Theorem~\ref{Theorem 3.2} has been proved.

To prove the second part of Theorem~\ref{Theorem 3.2}, fix the
electron number $N_{\rm e} > N_{\rm d}$, and assume that the
ground state energy is still $N_{\rm e}{\varepsilon }_0$.
Then the condition (\ref{f=0}), along with the fact that
$\abs{A_\uparrow}+\abs{A_\downarrow}>\abs{V}$, immediately
leads to $\Phi=0$.

%%%%%%%%%%%%%%%%%%%%%%%%%%%%%%%%%%%%%%%%
%%%%%%%%%%%%%%%%%%%%%%%%%%%%%%%%%%%%%%%%
\Section{Percolation representation}
\subsection{Representation}
In the present section, we shall prove Theorems~\ref{Theorem 2.3}
and \ref{Theorem 2.4}
concerning the behavior of grand canonical average of various
quantities.
The proofs are based on the complete characterization of the ground
states obtained in Section 3 and a percolation analysis.

Throughout the present section, we shall assume for simplicity that
the index set $V$ forms a $d$-dimensional hypercubic lattice (where
$d\ge2$) with periodic boundary conditions, and two sites $u$
and $v$ are directly connected with each other if and only if
$\abs{u-v}=1$.
This is precisely the case in the model of Section~2.
Generalizations to other models are straightforward.

The strategy to make use of the percolation type analysis in the
present context follows that of Mielke's work on the Hubbard
models on two dimensional line graphs \cite{Mielke2}.
Here we shall make the ideas presented in \cite{Mielke2} into a
rigorous proof, by presenting careful analysis of a non-independent
percolation problem.
(Unfortunately, the properties of the independent percolation.
As a consequence, the critical filling factors predicted in
\cite{Mielke2} are incorrect.
We stress, however, that the main body of \cite{Mielke2} is
completely rigorous, and that the present method can be also
applied to the models
of \cite{Mielke2} to establish rigorously the existence of
ferromagnetic order in a finite range of filling factor.)

In the present subsection, we will derive representations for
various expectation values in terms of a (non-independent)
percolation problem, and give some heuristic discussions about
behavior of the system.

We recall from Section~3.1 that the space of (many-electron) ground
states of $H$ is spanned by
the states
\begin{equation}
{ \Phi }_{A,\{{m}_{k}\}} =\{ \prod_{
k=1}^{ n}  ({\hat{S}}_{k}^{-}{)}^{(|{C}_{k}|/2)-
{m}_{k}}\}{ \Phi }_{A\uparrow },
\label{4.1}
\end{equation}
where $A$ is an arbitrary sublattice of the hypercubic lattice $V$.
The sets ${C}_{1},\cdots ,{C}_{n}$ are the connected components of
$A$
(in the usual sense), and ${m}_{k}=-|{C}_{k}|/2, 1-|{C}_{k}|/2,\cdots,
|{C}_{k}|/2$.
We first realize the trace in the definition of the grand canonical
like average (\ref{2.8}) by using the complete set of ground states
(\ref{4.1}).
For an arbitrary operator $O$, let us define its matrix element
$o(B,\{n_k\};A,\{m_k\})$ by
\begin{equation}
O{ \Phi }_{A,\{{m}_{k}\}}= \Psi  +
\sum_{ B,\{{n}_{k}\}}
o(B,\{{n}_{k}\};A,\{{m}_{k}\}){ \Phi }_{B,\{{n}_{k}\}},
\label{4.2}
\end{equation}
where $\Psi$ is a state orthogonal to the eigenspace of $H$ with the
eigenvalue $0$.
Since the states (\ref{4.1}) form a complete basis
of the eigenspace, the trace can be realized as
\begin{equation}
{\rm Tr} [O{P}_0]= \sum_{
A,\{{m}_{k}\}}  o(A,\{{m}_{k}\};A,\{{m}_{k}\}),
\label{4.3}
\end{equation}
where the sum is over all the subset possible $\{{m}_{k}\}$.

{}From (\ref{4.1}) and the definitions (\ref{1.5}), (\ref{1.6}) and
(\ref{1.7}), we immediately get
\begin{equation}
\hat{N}_{\rm e}{ \Phi }_{A,\{{m}_{k}\}}=|A|{ \Phi
}_{A,\{{m}_{k}\}},
\label{4.4}
\end{equation}
and
\begin{equation}
({S}_{\rm tot}^{(3)}{)}^{2}{ \Phi }_{A,\{{m}_{k}\}}=(
\sum_{ k=1}^{ n} { m}_{
k} {)}^{2}{ \Phi }_{A,\{{m}_{k}\}},
\label{4.5}
\end{equation}
where ${S}_{\rm tot}^{(3)}={ \sum}_{x\in  \Lambda }
{S}_{x}^{(3)}$.

For a fixed subset $A$, we define the corresponding subtrace by
\begin{equation}
{\rm Tr}_{A}[O]= \sum_{ \{{m}_{k}\}}
 o(A,\{{m}_{k}\};A,\{{m}_{k}\}).
\label{4.6}
\end{equation}
Straightforward calculations show
\begin{eqnarray}
{\rm Tr}_{A} [\exp (\mu  \hat{N}_{\rm e})]
&=& \sum_{{ m}_{ 1} =-
|{C}_{1}|/2}^{ |{C}_{1}|/2}  \cdots
\sum_{{ m}_{ n} =-
|{C}_{n}|/2}^{ |{C}_{n}|/2}  \exp (\mu  |A|)
\ret
&=& \prod_{ k=1}^{ n} {e}^{\mu  |{C}_{k}|}
(|{C}_{k}|+1)=W(A),
\label{4.7}
\end{eqnarray}
and
\begin{eqnarray}
{\rm Tr}_{A} [({ S}_{\rm tot}^{(3)}
{)}^{2} \exp (\mu  \hat{N}_{\rm e})]
&=&
\sum_{{ m}_{ 1}}  \cdots
 \sum_{{ m}_{ n}}  (
\sum_{ k=1}^{ n} { m}_{ k}
{)}^{2}\exp (\mu  |A|)
\ret
&=&W(A)\frac{1}{3}\sum_{k=1}^n
\frac{|{C}_{k}|}{2} \left(\frac{|{C}_{k}|}{2}+1\right).
\label{4.8}
\end{eqnarray}
{}From the definition (\ref{2.8}) of the average, and the $SU(2)$
invariance,
we get
\begin{equation}
\langle  ({\bf  S}_{\rm tot})^2 {\rangle
}_{\mu }={\frac{1}{Z}} \sum_{ A\subset
V}  W(A) \sum_{ k=1}^{ n} {\frac{
|{C}_{k}|}{2}} \left(\frac{|{C}_{k}|}{2}+1\right),
\label{4.9}
\end{equation}
where the ``partition function'' is given by
\begin{equation}
Z= \sum_{ A\subset  V}
W(A).
\label{4.10}
\end{equation}
Similarly we have
\begin{equation}
\langle  \hat{N}_{\rm e}{\rangle }_{\mu }={\frac{1}{Z}}
\sum_{ A\subset  V}  W(A)
\sum_{ k=1}^{ n}  |{C}_{k}|.
\label{4.11}
\end{equation}
Note that the right hand sides of eqs.
(\ref{4.9}) and (\ref{4.11}) can be regarded as expectation values in
a
percolation system.
The set $A$ may be regarded as a configuration of occupied sites.
The probability that a configuration $A$ appears is proportional to
the statistical weight $W(A)$ defined in (\ref{4.7}).
Such identification allows us to develop, as in \cite{Mielke2}, a
geometric picture for the
ferromagnetic-paramagnetic phase transition observed in the
present Hubbard model.

When $\mu$ is negative and its absolute value is large (compared to
1), (\ref{4.7}) suggests that the probability to find a configuration
with
many occupied sites is relatively small.
This means that, among many ground states (\ref{4.1}), only those
with
low electron filling factor have main contributions to the grand
canonical average.
We expect that the percolation system is in its low density phase,
where all the connected clusters (which are ${C}_{1},\cdots
,{C}_{n}$ in our case) have uniformly bounded sizes with large
probability.
Then the representation (\ref{4.9}) implies that the square of the
total
spin is proportional to the system size, which is a characteristic
behavior of a paramagnetic phase.

When $\mu$ is positive and large, (\ref{4.7}) suggests that the
probability
to find a configuration with many occupied sites is large.
This indicates that the electron filling factor should be close to its
maximum value ${\rho }_0$.
When the dimension $d$ is not less than the lower critical dimension
of the percolation problem, which is two, we expect that the
system is in the percolating phase, where one finds an infinitely
large
connected cluster with probability one.
Such a cluster has a dominant contribution in the representation
(\ref{4.9}), and we see that the square of the total spin becomes
proportional to the square of the system size.
This is the de
Although the present percolation problem is not a simple
independent percolation, we can use suitable stochastic geometric
techniques to get meaningful bounds for physical quantities.
In the following sections, we will make the above heuristic picture
into rigorous proofs.

%%%%%%%%%%%%%%%%%%%%%%%%%%%%%%%%%%%%%%%%
\subsection{Universal bounds}
We shall prove elementary bounds which are valid for any values of
$\mu$.

Since $|{C}_{k}|\ge1$, we see
\begin{equation}
{\frac{|{C}_{k}|}{2}}\left({\frac{|{C}_{k}|}{2}}+1\right)
\ge {\frac{3}{4}}|{C}_{k}|.
\label{4.12}
\end{equation}
Substituting the bound into (\ref{4.9}), we immediately get
\begin{equation}
\langle  ({\bf  S}_{\rm tot})^2
\rangle_\mu\ge {\frac{3}{4}}\langle  \hat{N}_{\rm e}
\rangle_\mu,
\label{4.13}
\end{equation}
for any value of $\mu$.
This is the lower bound in (\ref{2.14}).

We shall prove the upper bound for the electron number in (\ref{2.9}).
Fix a site $u\in V$, and let $A$ be an arbitrary configuration in
which $u$ is occupied.
By $A^*$ we denote the unique configuration obtained from $A$ by
eliminating the site $u$.
Then the weights for two subsets satisfy
\begin{equation}
{\frac{W(A^*)}{W(A)}}={\frac{{ \prod}_{j}
(|{C'}_{j}|+1)}{{e}^{\mu }(|C|+1)}},
\label{4.14}
\end{equation}
where $C$ is the connected cluster in $A$ which contains $u$, and
${C'}_{j}$ are the connected components of $C\backslash \{u\}$.
Noting that $|C|=1+{ \sum}_{j} |{C'}_{j}|$ , the above ratio can
be bounded as
\begin{equation}
{\frac{W(A^*)}{W(A)}}\ge {\frac{1+{ \sum}_{j}
|{C'}_{j}|}{{e}^{\mu }(2+{ \sum}_{j} |{C'}_{j}|)}}\ge
{\frac{1}{2{e}^{\mu }}}.
\label{4.15}
\end{equation}

For a general event concerning a configuration $A$, we define
\begin{equation}
{Z}_{\rm event} = \sum_{ A\subset  V}
 W(A)\chi[\mbox{event}],
\label{4.16}
\end{equation}
where $\chi$ is the indicator fu$\chi[\mbox{True}] =1$,
$\chi[\mbox{False}] = 0$.
The bound (\ref{4.15}) implies that
\begin{eqnarray}
{Z}_{u\in  A} &=& \sum_{A\subset V} W(A) \chi[u\in A]
\le 2{e}^{\mu }
\sum_{A\subset V}  W(A^*) \chi[u\in A]
\ret
&=& 2e^\mu \sum_{A'\subset V} W(A') \chi[u\not\in A']
=2{e}^{\mu }{Z}_{u\not\in  A},
\label{4.17}
\end{eqnarray}
where we have used the fact that, when $A$ runs over all the
configurations with $u\in A$, $A^*$ runs over all the configurations
with $u\not\in A^*$.
This immediately implies
\begin{equation}
\langle  \chi  [u\in  A]{\rangle }_{\mu }={\frac{{Z}_{u\in
 A}}{{Z}_{u\in  A}+{Z}_{u\not\in  A}}}\le {\frac{1}{1+{e}^{-
\mu }/2}}.
\label{4.18}
\end{equation}
By summing up the inequality over $u$, we get
\begin{equation}
\langle  \hat{N}_{\rm e}{\rangle }_{\mu }\le {\frac{|V|}{1+{e}^{-\mu
}/2}},
\label{4.19}
\end{equation}
which is nothing but the desired upper bound in (\ref{2.9}).

We shall now prove the lower bound in (\ref{2.9}).
Let $b$ be a $3\times\cdots\times3$ (hyper-)cubic region in $V$,
and $u$ be the site at the center of $b$.
Take an arbitrary configuration $A$ with $A\cap  b=\emptyset $,
and let $A^*$ be the unique configuration obtained by adding the
site $u$ to $A$.
Then we have $2{e}^{\mu }W(A)=W(A^*)$.
As in the above we see that
\begin{equation}
\langle  \chi  [A\cap  b\ne \emptyset  ]{\rangle }_{\mu
}={\frac{{Z}_{A\cap  b\ne \emptyset }}{{Z}_{A\cap  b=\emptyset
}+{Z}_{A\cap  b\ne \emptyset }}}\ge {\frac{{Z}_{A\cap
b=u}}{{Z}_{A\cap  b=\emptyset }+{Z}_{A\cap
b=u}}}=\frac{1}{1+e^{-\mu}/2}.
\label{4.20}
\end{equation}
The quantity in the left hand side is the probability that there is at
least one occupied site among the $3^d$ sites in $b$.
By summing up (\ref{4.20}) over all the nonoverlapping
$3\times\cdots\times3$ (hyper)cubes in $V$, we get
\begin{equation}
\langle  \hat{N}_{\rm e}{\rangle }_{{\frac{|V|}{{3}^{d}(1+{e}^{-\mu }/2)}},
\label{4.21}
\end{equation}
which is the desired lower bound in (\ref{2.9}).

%%%%%%%%%%%%%%%%%%%%%%%%%%%%%%%%%%%%%%%%
\subsection{Low density bounds}
We shall prove the bounds which are valid when the chemical
potential $\mu$ is negative and its absolute value is large.

Let $\setC [A] = \{{C}_{1},\cdots ,{C}_{n}\}$ be the set of
connected clusters in a configuration $A$.
Take a connected set $C$ of sites in $V$.
For any configuration $A$ with $C\in  \setC [A]$, we denote
by $A^*$ the configuration obtained by eliminating $C$ from $A$.
When $A$ runs over all the configurations with $C\in\setC[A]$,
$A^*$ runs over all the configurations with
$\bar{C}\cap A^*=\emptyset$, where $\bar{C}$ is a set obtained by
adding neighboring sites to $C$.
We have $W(A)={e}^{\mu  |C|}(|C|+1)W(A^*)$, and thus
\begin{equation}
\langle  \chi  [C\in \setC [A]]{\rangle }_\mu
= \frac{Z_{C\in \setC [A]}}
{Z_{\bar{C}\cap A\ne\emptyset}+Z_{\bar{C}\cap A=\emptyset}}
\le
\frac{Z_{C\in \setC [A]}}{Z_{\bar{C}\cap A=\emptyset}}
={e}^{\mu |C|}(|C|+1).
\label{4.22}
\end{equation}
Note that, when $\mu$ is negative and large, the right hand side
converges to zero rapidly as the cluster becomes large.
{}From the representations (\ref{4.9}) and (\ref{4.11}), we get
\begin{eqnarray}
\langle  ({\bf S}_{\rm tot}{)}^{2} {
\rangle }_{\mu }-{\frac{3}{4}}\langle  \hat{N}_{\rm e}{ \rangle
}_{\mu }
&=&
 \sum_{ C \subset  V}
{\frac{ 1}{4}} ({|C|}^{2}-|C|)\langle  \chi  [C\in
\setC[A]]{ \rangle }_{\mu }
\ret
&=&
|V| \sum_{ C\ni  o} {\frac{ 1}{4}}
({|C|}^{2}-|C|)\langle  \chi  [C \in \setC[A]]{ \rangle
}_{\mu },
\label{4.23}
\end{eqnarray}
where, in the right hand side, the sum is over all the connected set
$C$ which contains a fixed lattice site $o$.
We have made use of the translation invariance to get the final
equality.
The standard argument showith $n$ sites can bounded from above by ${a}^{n}$
where $a$ is a
positive constant which depends only on the dimension $d$.
By substituting the bound (\ref{4.22}) into (\ref{4.23}), we get
\begin{eqnarray}
\langle({\bf S}_{\rm tot})^{2}
{ \rangle }_{\mu }-{\frac{3}{4}}\langle  \hat{N}_{\rm e}{
\rangle }_{\mu }
&\le & |V| \sum_{n=1}^{
\infty } a^n  {\frac{1}{4}}({n}^{2}-n)(n+1){e}^{\mu  n}
\ret
&\le & {c}_{3}\abs{V}{e}^{2\mu },
\label{4.24}
\end{eqnarray}
where the final bound is valid when $-\mu$ is sufficiently large and
the sum converges.
Here ${c}_{3}$ is a positive finite constant which depends only on
the
dimension $d$.
The upper bound in (\ref{2.14}) has been proved.

%%%%%%%%%%%%%%%%%%%%%%%%%%%%%%%%%%%%%%%%
\subsection{High density bounds}
We shall prove the bounds which are valid when $\mu$ is positive
and large enough.
Our proofs are based on a variation of the Peierls
argument, which is standard in spin systems \cite{Griffiths} and
percolation \cite{Grimmet}.
Let us recall that
we have assumed that
$V$ is a $d$-dimensional hypercubic lattice with periodic
boundary conditions.

Let $\overline{A} = V\backslash A$, which may be called the set of
unoccupied sites, or defects.
Again this can be decomposed into connected components as
$\overline{A}={D}_{1}\cup  \cdots \cup  {D}_{m}$, and we
denote by $\setD [A]$ the set $\{{D}_{1},\cdots ,{D}_{m}\}$.
Take a connected set $D$.
The set $V\backslash D$ decouples into several connected
components.
We call the largest connected component the exterior of $D$, and
other components the interior of $D$.

Take a connected set $D$, and let $A$ be an arbitrary configuration
with $D\in \setD [A]$.
We denote by
$A^*$ the configuration obtained by adding to $A$ all the sites of
$D$.
Let $C^D_j\in \setC[A]$ with $j = 0, \cdots, p $ be the connected
clusters in $A$ whicWe choose the numbering so that $C^D_0$ lies in the
exterior of $D$,
and $C^D_j$ with $j=1,\cdots,p$ in the interior.
See Fig.2.

Now the ratio of the weights for the configurations $A$ and $A^*$
can be evaluated as
\begin{eqnarray}
{\frac{W(A)}{W(A^*)}}
&=&
{e}^{-\mu  |D|}{\frac{
 \prod_{ j=0}^{ p}
(|C^D_j|+1)}{( \sum_{ j=0}^{ p}
 |C^D_j|+|D|+1)}}
\ret
&\le &
 {e}^{-\mu  |D|} \prod_{ j=1}^{
 p}  (|C^D_j|+1).
\label{4.25}
\end{eqnarray}
Since each $C^D_j$ with $j = 1,2,\cdots, p$ is in the interior of
$D$, it must be surrounded by sites in $D$.
Let ${d}_{j}$ be the number of sites in $D$ which are directly
surrounding the cluster $C^D_j$.
Then we get the following bound.
\begin{eqnarray}
\prod_{ j=1}^{ p}
 (|C^D_j|+1)
&\le &
 c' \max_{
\{{d}_{1},\cdots ,{d}_{p}\},{d}_{j}\ge 1,{ \sum}_{j=1}^{p}
{d}_{j}\le 2|D|} \prod_{ j=1}^{ p}
({d}_{j}{)}^{d/(d-1)}
\ret
&\le &
 c'({\frac{|D|}{p}}{)}^{pd/(d-1)}\le c' \max_q
 ({\frac{|D|}{q}}{)}^{qd/(d-1)}
\ret
&\le & c' \exp [\alpha  |D|],
\label{4.26}
\end{eqnarray}
where $c'$ is a constant, and $\alpha=d/\{e(d-1)\}$.
Thus we get the bound
\begin{equation}
\langle  \chi  [D\in  \setD [A]]{\rangle }_{\mu }\le
{\frac{{Z}_{D\in  \setD [A]}}{{Z}_{D\not\in  \setD
[A]}}}\le c'{e}^{-(\mu  -\alpha  )|D|}.
\label{4.27}
\end{equation}
Note that when $\mu$ is sufficiently large, the probability to find a
larger defect becomes exponentially small.

We will construct a lower bound for the electron number.
{}From the representation (\ref{4.11}), and the definition of the
defects
$\{{D}_{1},\cdots ,{D}_{m}\}$, we get
\begin{eqnarray}
|V|-\langle  \hat{N}_{\rm e}{ \rangle }_{\mu
}
&=&
{\frac{1}{Z}} \sum_{ A \subset
 V}  W(A) \sum_{ k=1}^m
|{D}_{k}|= \sum_{ D \subset
V}  \langle  \chi  [D\in  \setD [A]]{ \rangle
}_{\mu }|D|
\ret
&=&
|V| \sum_{ D\ni  o}  \langle
\chi  [D \in   \setD [A]]{ \rangle }}|D|,
\label{4.28}
\end{eqnarray}
where the sum in the right hand side is over all the connected sets
which contains a fixed lattice site $o$.
We have made use of the translation invariance as in (\ref{4.23}).
{}From the bound (\ref{4.27}) and the same entropy estimate as in
(\ref{4.24}), we
get
\begin{equation}
|V|-\langle  \hat{N}_{\rm e}{\rangle }_{\mu }
\le
|V| \sum_{ n=1}^{ \infty } {
a}^{ n} nc'{e}^{-(\mu  -\alpha  )n}
\le  {c}_{1}|V|{e}^{-\mu },
\label{4.29}
\end{equation}
where the final inequality holds when $\mu$ is large enough and the
sum converges, and ${c}_{1}$ is a positive finite constant which
depends only on the dimension.
Thus we have proved the lower bound (\ref{2.10}) for the electron
number.

We will now prove the lower bound for the total spin in (\ref{2.13}),
which
establishes the appearance of ferromagnetism.
(Note that the upper bound in (\ref{2.13}) is trivial.)  Let $A$ be a
configuration and $\setD[A]$ be the set of the corresponding defects.
By ${\rm ext}(D)$ we denote the exterior of an deffect $D$.
Define
\begin{equation}
{C}_{\rm ext}=\bigcap_{D \in \setD [A]} {\rm ext}(D),
\label{4.30}
\end{equation}
which is indeed an element in $\setC[A]$.
For a connected cluster (defect) $D$, we denote by $v(D)$ the total
number of sites in $D$ and in its interior.
Then there is a trivial inequality
\begin{equation}
|{C}_{\rm ext}|\ge |V|- \sum_{ D\in
\setD [A]}  v(D).
\label{4.31}
\end{equation}
{}From (\ref{4.31}) and the representation (\ref{4.9}), we now have
\begin{eqnarray}
\langle  ({\bf  S}_{\rm tot}{)}^{2}
 { \rangle }_{\mu }
&\ge &
 {\frac{1}{Z}}
\sum_{ A \subset  V}
W(A){\frac{|{C}_{\rm ext}|}{2}}
\left({\frac{|{C}_{\rm ext}|}{2}}+1\right)
\ret
&\ge &
 {\frac{1}{Z}} \sum_{ A \subset
 V}  W(A){\frac{|V|-{ \sum}
v(D)}{2}}\left({\frac{|V|-{ \sum} v(D)}{2}}+1\right)
\ret
&\ge &
{\frac{1}{Z}} \sum_{ A \subset
 V}  W(\left(|V|+{\frac{1}{2}}\right) \sum_{ D \in
  \setD [A]}  v(D)\right\}
\ret
&=&
{S}_{ {\rm max}}({S}_{ {\rm max}}+1)-
\left(|V|+{\frac{1}{2}}\right)|V|
\sum_{ D\ni  o}  \langle  \chi
[D \in   \setD [A]]{\rangle }_{\mu }v(D),
\label{4.32}
\end{eqnarray}
where ${S}_{{\rm max}}=|V|/2$.
The sum in the most right hand side is over all the connected
clusters including a fixed lattice site $o$.
We have again made use of the translation invariance as in
(\ref{4.23}) and
(\ref{4.28}).
Substituting the bounds $v(D)\le {|D|}^{d/(d-1)}$ and the entropy
estimate (as in (\ref{4.24})) into (\ref{4.32}), we finally get
\begin{eqnarray}
\langle  ({\bf  S}_{\rm tot}{)}^{2}
 {\rangle }_{\mu }
&\ge &
{S}_{ {\rm max}}({S}_{ {\rm max}}+1)-
(|V|+{\frac{1}{2}})|V| \sum_{ n=1}^{
\infty } { a}^{ n} {n}^{d/(d-1)}{e}^{-(\mu
 -\alpha  )n}
\ret
&\ge &
{S}_{ {\rm max}}({S}_{ {\rm max}}+1)\{1-{c}_{2}{e}^{-\mu
}\},
\label{4.33}
\end{eqnarray}
for sufficiently large values of $\mu$.
Here ${c}_{2}$ is a positive finite constant which depends only on
the
dimension $d$.
Thus the lower bound in (\ref{2.13}) is proved.

%%%%%%%%%%%%%%%%%%%%%%%%%%%%%%%%%%%%%%%%
%%%%%%%%%%%%%%%%%%%%%%%%%%%%%%%%%%%%%%%%
\Section{Correlation functions}
In the present section, we evaluate electron-electron
correlation functions for the model of Section~2 with
$N_{\rm d}-2d+1 \le N_{\rm e} \le N_{\rm d}$.
As we have mentioned in Section~2.3, the result implies that the
coherence length in the ground states is $\lambda^{-1}$.

Let $\microage{\, \cdots \,}$ be the usual microcanonical average
for some fixed $N_{\rm e}\leq N_{\rm d}$. For an arbitrary operator
$O$,
it may be expressed as
\begin{equation}
\microage{O} = \frac{\sum_{
A,\{{m}_{k}\}\,;\,|A|=N_{\rm e}}  o(A,\{{m}_{k}\};A,\{{m}_{k}\})}
{\sum_{A,\{{m}_{k}\}\,;\,|A|=N_{\rm e}}1}.
\label{mavO}
\end{equationThen one obtains

\begin{lemma}
If for a given $N_{\rm e}\le N_{\rm d}$ all ground states have the
total spin
$S_{\rm tot}=N_{\rm e}/2$, then
\begin{equation}
\sum_{\sigma}\microage{c^\dagger_{x\sigma} c_{y\sigma}}
= \frac{N_{\rm e}}{N_{\rm d}}
\sum_{u\in V} \kappa_x^{(u)}\varphi_y^{(u)}.
\label{correl1}
\end{equation}
\label{correl lemma}
\end{lemma}
\begin{proof}{Proof}
Since the operator
$O_{xy}=\sum_\sigma c^\dagger_{x\sigma}c_{y\sigma}$
is $SU(2)$ invariant, it suffices to consider the expectation value in
the subspace where all spins are $\uparrow$.
We thus get
\begin{equation}
\microage{O_{xy}}={N_{\rm d} \choose N_{\rm e}}^{-1}
\sum_{
A\,;\,|A|=N_{\rm e}}  o_{xy}(A,\uparrow;A,\uparrow),
\label{correl2}
\end{equation}
where
$o_{xy}(A,\uparrow;A,\uparrow)$
are the matrix elements of $O_{xy}$ in the subspace with all spins
$\uparrow$.
Using the expansion formulas (\ref{expcdag}) and (\ref{expc}),
these matrix elements may be written as
\begin{eqnarray}
o_{xy}(A,\uparrow;A,\uparrow)
&=&\sum_{u,v\in V} \kappa_x^{(u)}\varphi_y^{(v)}
\tilde o_{uv}(A,\uparrow;A,\uparrow)
\ret
&=&\sum_{u\in V} \kappa_x^{(u)}\varphi_y^{(u)}
\tilde o_{uu}(A,\uparrow;A,\uparrow),
\label{correl3}
\end{eqnarray}
where
$\tilde o_{uv}(A,\uparrow;A,\uparrow)$
are the matrix elements of
$\tilde O_{uv}=a^\dagger_{u\uparrow} b_{v\uparrow}$
and we used the relation
\begin{equation}
\tilde o_{uv}(A,\uparrow;A,\uparrow)
=\delta_{uv}
\tilde o_{uu}(A,\uparrow;A,\uparrow).
\label{correl4}
\end{equation}
Now,
\begin{eqnarray}
\sum_{
A\,;\,|A|=N_{\rm e}}
\tilde o_{uv}(A,\uparrow;A,\uparrow)
&=&\sum_{A\ni u, |A|=N_{\rm e}}1
\ret
&=&{N_{\rm d}-1 \choose N_{\rm e}-1},
\end{eqnarray}
which yields the desired result.
\end{proof}

Clearly, Lemma \ref{correl lemma} may be applied in the case
$N_{\rm e}=N_{\rm d}$ if $V$ is connected,
and in the case $N_{\rm e}=1$,
which is trivial. As a
\begin{coro}
Consider the model of Section~2 in the subspace with the electron
number fixed to some value
$N_{\rm d}-2d+1 \le N_{\rm e} \le N_{\rm d}$. Then
\begin{equation}
\sum_{\sigma}
\microage{c^\dagger_{x\sigma} c_{y\sigma}}
\simeq
\sqrt{\frac{\pi}{2}}(2\pi)^{-d}\lambda^{(d-3)/2}
(-1)^{\norm{x-y}_1}\abs{x-y}^{-(d-1)/2}\exp[-\lambda\abs{x-y}]
\label{2.16}
\end{equation}
for $L\gg1$ and $\lambda\abs{x-y}\gg1$,
where $\norm{x}_1=\sum_{j=1}^d|x_j|$.
\end{coro}
\begin{proof}{Proof}
Since any $A\subset V$ with $|V|-2d+1\le|A|$ is connected,
we find that any ground state has $S_{\rm tot}=N_{\rm e}/2$
if $N_{\rm d}-2d+1 \le N_{\rm e} \le N_{\rm d}$.
This allows us to use Lemma~\ref{correl lemma}.

It remains to calculate $\kappa_x^{(y)}$.
The Gramm matrix (\ref{Guv}) for the basis (\ref{2.7}) is given by
\begin{equation}
(G)_{uv} = \left\{
\begin{array}{ll}
1+2d\lambda^{-2} & \mbox{if $u=v$},\\
\lambda^{-2} & \mbox{if $\abs{u-v}=1$},\\
0 & \mbox{otherwise}.
\end{array}
\right.
\end{equation}
The inverse can be calculated as
\begin{equation}
(G^{-1})_{uv} = (-1)^{\norm{u-v}_1}
(2\pi)^{-d} \int_{k\in(-\pi,\pi]^d}
d^dk\,\frac{\lambda^2 e^{ik\cdot(u-v)}}
{\sum_{j=1}^d(2\sin[k_j/2])^2 + \lambda^2}.
\label{Ginverse}
\end{equation}
The integral $(2\pi)^{-d} \int d^dk$ is a shorthand for the
sum $L^{-d}\sum_k$ where $k=(2\pi n_1/L,\ldots, 2\pi
n_d/L)$ with integers $n_j$ such that $-L/2<n_j\le L/2$.
Since the definition (\ref{kappa}) implies
$\kappa_x^{(y)} = (G^{-1})_{yx}$ for $x\in V$, the desired expression
for the correlation function (\ref{2.16}) follows by combining
(\ref{correl1}), (\ref{Ginverse}) and the standard asymptotic
evaluation of the integral (sum) in (\ref{Ginverse}).
\end{proof}

Just as one cannot conclude from the vanishing charge gap
(\ref{2.15}) in the case
$N_{\rm d}-2d+1 \le N_{\rm e} < N_{\rm d}$
that the systemdecay in (\ref{2.16}) that the electrons are localized. In the
case
$N_{\rm e} \le N_{\rm d}$
this exponential decay is due to the ground state degeneracy.
$\lambda ^{-1}$ is the phase coherence length and not the
localization length.

Before closing the section, let us briefly discuss why we are
 not able to calculate the correlation function in general
situation with $\mu<\infty$.
In the grand canonical ensemble, we also have
\begin{eqnarray}
\average{c^\dagger_{x\sigma} c_{y\sigma}}
&=& \sum_{u,v\in V} \kappa_x^{(u)}\varphi_y^{(v)}
\average{a^\dagger_{u\sigma} b_{v\sigma}}
+\sum_{u\in V} \kappa_x^{(u)}
\average{a^\dagger_{u\sigma} d_{y\sigma}}
+\sum_{u\in V} \varphi_y^{(u)}
\average{d^\dagger_{x\sigma} b_{u\sigma}}
+\average{d^\dagger_{x\sigma}d_{y\sigma}}
\ret
&=& \sum_{u,v\in V} \kappa_x^{(u)}\varphi_y^{(v)}
\average{a^\dagger_{u\sigma} b_{v\sigma}}.
\end{eqnarray}
But (\ref{correl4}) only holds in the subspace with
$S_{\rm tot}=N_{\rm e}/2$.
The reason is that only in this subspace all the states that can
be constructed using the operators $a^\dagger_{u\sigma}$ are ground
states of $H$. Therefore, the dual basis in this subspace is simply
given by all the states constructed using the operators
$b_{u\sigma}$, and
$\tilde o_{uv}(A,\uparrow;A,\uparrow)$
has the simple form (\ref{correl4}). In all the other subspaces
with $S_{\rm tot}<N_{\rm e}/2$ the dual basis to the states
(\ref{3.6}) cannot
easily be constructed.

The difficulty for the models with $\mu<\infty$ may be
expected from a physical point of view.
The ground states for $N_{\rm e}=N_{\rm d}$ are essentially
the same as that of a noninteracting spinless fermion system.
Calculating correlation functions should be of no difficulty.
When $\mu<\infty$, on the other hand, the ground states
reflect strong correlation effect, and have qstructure.
Though the expression for the ground state (\ref{3.6})
appears rather simple in the nonorthogonal basis, it will
become highly complicated when expressed in the standard
orthogonal basis.
%%%%%%%%%%%%%%%%%%%%%%%%%%%%%%%%%%%%%%%%
%%%%%%%%%%%%%%%%%%%%%%%%%%%%%%%%%%%%%%%%
\Section{Spin Hamiltonian and spin wave excitations}
In the present section, we rigorously derive a ``spin Hamiltonian'' of
our Hubbard model.
More precisely we show that the interaction Hamiltonian
$H_{\rm int}$, when projected onto a certain subspace which
includes ground states, exactly reduces to the Hamiltonian of the
ferromagnetic Heisenberg model.
In the model with exactly half-filled degenerate band, we prove that
there is a set of low energy excited states whose excitation
energies are bounded from above by that of the spin wave
excitations of the Heisenberg model.
This result suggests that our Hubbard model has a ``normal'' spin
excitation structure.
Based on a trial state calculation, Kusakabe and Aoki recently
\cite{Kusakabe2} argued that the closely related Hubbard model with
$U=\infty$ has low-lying excitations which are expected for a
``normal'' itinerant electron system with ferromagnetic order.

We first discuss the general class of models as in Section~3.
For a subset $A\subset V$, we define a (non-orthogonal) projection
operator $P_A$ by the following procedure.
Given an arbitrary state $\Phi$, we can uniquely decompose it as
\begin{equation}
\Phi = \Psi+\sum_{A_\uparrow,A_\downarrow}
f(A_\uparrow,A_\downarrow)
\prod_{u\in A_\uparrow} a^\dagger_{u\uparrow}
\prod_{v\in A_\downarrow} a^\dagger_{v\downarrow}\Phi_0,
\end{equation}
where $\Psi$ is orthogonal to the space with
$H_{\rm hop}\Phi=N_{\rm e}\varepsilon_0\Phi$.
We define
\begin{equation}
P_A\Phi = \sum_{A_\uparrow,A_\downarrow;\,
A_\uparrow\cup A_\downarrow=A,\,
A_\uparrow\cap A_\f(A_\uparrow,A_\downarrow)
\prod_{u\in A_\uparrow} a^\dagger_{u\uparrow}
\prod_{v\in A_\downarrow} a^\dagger_{v\downarrow}\Phi_0.
\end{equation}

By using the expansion formulas (\ref{expcdag}) and (\ref{expc}), we
get
\begin{equation}
n_{x\uparrow} n_{x\downarrow}
= (\sum_q
\kappa_x^{(q)}a^\dagger_{q\uparrow}+d^\dagger_{x\uparrow})
(\sum_r \varphi_x^{(r)}b_{r\uparrow}+d_{x\uparrow})
(\sum_s
\kappa_x^{(s)}a^\dagger_{s\downarrow}+d^\dagger_{x\downarrow})
(\sum_t \varphi_x^{(t)}b_{t\downarrow}+d_{x\downarrow}).
\label{nn}
\end{equation}
Let a site $x\in\Lambda$ be such that there is an index $u\in V$
with
$x(u)=x$, where $x(u)$ is defined in the {\em quasi locality\/}
condition B1 in Section~3.1.
Then from (\ref{nn}) we get
\begin{equation}
P_A n_{x\uparrow} n_{x\downarrow} P_A
= P_A \sum_{q,s\in V}
\kappa_x^{(q)}\varphi_x^{(u)}\kappa_x^{(s)}\varphi_x^{(u)}
a^\dagger_{q\uparrow}b_{u\uparrow}
a^\dagger_{s\downarrow}
b_{u\downarrow} P_A
= 0.
\end{equation}
Next take a site $x\in\Lambda$ such that there are indices $u,v\in
V$ with
$\varphi_x^{(u)}\varphi_x^{(v)}\ne0$.
{}From the {\em local connectivity\/} condition B2 and (\ref{nn}), we
have
\begin{eqnarray}
P_A n_{x\uparrow} n_{x\downarrow} P_A
&=& P_A \sum_{q,r,s,t\in V}
\kappa_x^{(q)}\varphi_x^{(r)}\kappa_x^{(s)}\varphi_x^{(t)}
a^\dagger_{q\uparrow}b_{r\uparrow}
a^\dagger_{s\downarrow}b_{t\downarrow} P_A
\ret
&=& \kappa_x^{(u)}\kappa_x^{(v)}\varphi_x^{(u)}\varphi_x^{(v)}
(a^\dagger_{u\uparrow}b_{u\uparrow}
a^\dagger_{v\downarrow}b_{v\downarrow} +
a^\dagger_{u\uparrow}b_{v\uparrow}
a^\dagger_{v\downarrow}b_{u\downarrow}
\ret
&&+
a^\dagger_{v\uparrow}b_{u\uparrow}
a^\dagger_{u\downarrow}b_{v\downarrow} +
a^\dagger_{v\uparrow}b_{v\uparrow}
a^\dagger_{u\downarrow}b_{u\downarrow})P_A.
\label{nn2}
\end{eqnarray}
Let us introduce spin operators for the state $\varphi^{(u)}$ by
\begin{eq\tilde{S}_u^{(\alpha)}=\sum_{\sigma,\tau=\uparrow,\downarrow}
a^\dagger_{u\sigma}p_{\sigma\tau}^{(\alpha)}b_{u\tau}/2,
\end{equation}
where $p_{\sigma\tau}^{(\alpha)}$ with $\alpha = 1,2,3$ are the
Pauli matrices (\ref{Pauli}).
The operators $\tilde{S}_u^{(\alpha)}$ are not self-adjoint, but have
exactly the same algebraic structure ({\em i.e.\/}, commutation
relations) as
the standard spin operators.
Using these operators, (\ref{nn2}) can be rewritten as
\begin{equation}
P_A n_{x\uparrow} n_{x\downarrow} P_A
=\kappa_x^{(u)}\kappa_x^{(v)}\varphi_x^{(u)}\varphi_x^{(v)}
\left(\frac{1}{2}-2\tilde{\bf S}_u\cdot\tilde{\bf S}_v\right)
P_A.
\label{nn3}
\end{equation}

By substituting (\ref{nn3}) into the definition (\ref{1.3}) of the
interaction Hamiltonian, we get the desired ``spin Hamiltonian''
\begin{equation}
P_A H_{\rm int} P_A
= \sum_{u,v\in V} J_{uv}
\left(\frac{1}{2}-
2\tilde{\bf S}_u\cdot\tilde{\bf S}_v\right) P_A,
\label{3.18}
\end{equation}
where the ``exchange interaction'' is given by
\begin{equation}
J_{uv} = \sum_x U_x
\kappa_x^{(u)}\kappa_x^{(v)}\varphi_x^{(u)}\varphi_x^{(v)}.
\end{equation}
If the ``exchange interaction'' $J_{uv}$ is nonnegative, the
``spin Hamiltonian'' (\ref{3.18}) is nothing but the Hamiltonian of the
ferromagnetic Heisenberg model.
We expect that the ``spin Hamiltonian'' (\ref{3.18}) approximately
describes low
energy excited states which involve only spin degrees of freedom.

In \cite{Tasaki2} it was shown that a class of models with certain
global symmetry has $J_{uv}>0$ for directly connected $u$, $v$.
The proof of ferromagnetism in \cite{Tasaki2} was based on this
observation.
Note, however, that the proof in Section~3.2 no longer requires
symmetry conditions.

We can explicitly evaluate $J_{uv}$ for the model of
Section~2.
By combining the definition (\ref{2.7}) of the basis state
$\varphi_x{(u)}$, the dand the
formula (\ref{Ginverse}) for $(G^{-1})_{uv}$, we get
\begin{equation}
J_{uv} = \left\{\begin{array}{ll}
J(\lambda) > 0 & \mbox{if $\abs{u-v}=1$},
\\
0 & \mbox{otherwise},
\end{array}\right.
\end{equation}
where
\begin{eqnarray}
J(\lambda) &=& U'
\left((2\pi)^{-d}\int_{k\in(-\pi,\pi]^d} d^dk
\frac{1-\cos(k_1)}
{\sum_{j=1}^d(2\sin(k_j/2))^2 + \lambda^2}\right)^2
\ret
&\simeq&\left\{\begin{array}{ll}
U' C_d & \mbox{as $\lambda\downarrow0$},
\\
U' \lambda^{-4} & \mbox{for $\lambda\gg1$}.
\end{array}\right.
\end{eqnarray}
The constant $C_d$ depends only on the dimension $d$.

The form of the ``spin Hamiltonian'' (\ref{3.18}) suggests that the
model has spin wave excitations when there is a ferromagnetic
order in the ground states.
For the model with $N_{\rm e}=N_{\rm d}$, we can prove the
existence of low energy excited states as follows.

\begin{theorem}
Consider the model of Section~2 in the subspace with the electron
number fixed to $N_{\rm e}=N_{\rm d}$.
To each wave vector $k=(2\pi n_1/L,\ldots,2\pi
n_d/L)\ne(0,\ldots,0)$ with integers $-L/2<n_j\le L/2$,
there corresponds an excited state whose excitation energy is
bounded from above by a function $E_L(k)$.
In the infinite volume limit $L\to\infty$, $E_L(k)$ converges to an
analytic function $E(k)$ which satisfies
\begin{equation}
E(k)=J(\lambda)|k|^2+O(|k|^4).
\label{disp}
\end{equation}
\end{theorem}

Recall that the dispersion relation of the spin wave excitations
for the Heisenberg Hamiltonian (\ref{3.18}) is
\begin{equation}
E_{\rm SW}(k)=2J(\lambda)\sum_{j=1}^d(1-\cos k^{(j)}),
\end{equation}
which has the same asymptotic behavior as (\ref{disp}).

\bigskip
\begin{proof}{Proof}
Let us define the ``one magnon state'' by
\begin{equation}
\Psi_k = \sum_{u\in V} e^{ik\cdot u}
\tilde{S}_u^- \Phi_{V\uparrow},
\end{equation}
where $\tilde{S}_u^-=\a^\dagger_{u\downarrow}b_{u\uparrow}$, and
$\Phi_{V\uparrow}=\prod_{v\in V} a^\dagger_{v\uparrow} \Phi_0$.
Note that $\Psi_k$ with different $k$ are orthogonal to each
other, and $\Psi_k$ with $k\ne(0,\ldots,0)$ is orthogonal to the
ground states.
If we denote the energy expectation value as
\begin{equation}
E_L(k)=\frac{(\Psi_k, H\,\Psi_k)}{(\Psi_k, \Psi_k)},
\label{ELk}
\end{equation}
the first part of the theorem follows from the standard variational
argument.

It now remains to evaluate $E_L(k)$.
As for the denominator of (\ref{ELk}), we have
\begin{eqnarray}
(\Psi_k,\Psi_k)&=&\sum_{u,v}e^{ik\cdot(v-u)}
(a^\dagger_{u\downarrow}b_{u\uparrow}\Phi_{V\uparrow},
a^\dagger_{v\downarrow}b_{v\uparrow}\Phi_{V\uparrow})
\ret
&=&\sum_{u,v}e^{ik\cdot(v-u)}(G)_{uv}(G^{-1})_{uv}
(\Phi_{V\uparrow},\Phi_{V\uparrow})
\ret
&=&L^d \sum_u e^{-ik\cdot u}(G)_{uo}(G^{-1})_{uo}
(\Phi_{V\uparrow},\Phi_{V\uparrow}),
\label{deno}
\end{eqnarray}
where $o$ denotes an arbitrary fixed site in $V$.
The second line of (\ref{deno}) follows from the anticommutation
relations
$\{a_{u\sigma},a^\dagger_{v\tau}\}=(G)_{uv}\delta_{\sigma\tau}$,
$\{b_{u\sigma},b^\dagger_{v\tau}\}=
(G^{-1})_{uv}\delta_{\sigma\tau}$,
and the relations $b^\dagger_{u\uparrow}\Phi_{V\uparrow}
=a_{u\downarrow}\Phi_{V\uparrow}=0$.
The third line follows from the translation invariance.

The numerator of (\ref{ELk}) can be evaluated in a similar manner,
although the calculation is more complicated.
By using the representation (\ref{nn}),
we have
\begin{eqnarray}
(\Psi_k,n_{x\uparrow}n_{x\downarrow}\Psi_k)
&=&\sum_{u,v}e^{ik\cdot(v-u)}
\sum_{q,r,s,t}
\kappa_x^{(q)}\varphi_x^{(r)}\kappa_x^{(s)}\varphi_x^{(t)}
(a^\dagger_{u\downarrow}b_{u\uparrow}\Phi_{V\uparrow},
a^\dagger_{q\uparrow}b_{r\uparrow}
a^\dagger_{s\downarrow}b_{t\downarrow}
a^\dagger_{v\downarrow}b_{v\uparrow}\Ph\ret
&=&\sum_{u,v}e^{ik\cdot(v-u)}\sum_{q,r,s,t}
\kappa_x^{(q)}\varphi_x^{(r)}\kappa_x^{(s)}\varphi_x^{(t)}
\delta_{vt}(G)_{us}
\{\delta_{qr}(G^{-1})_{uv}-\delta_{qv}(G^{-1})_{ur}\}
\ret&&\times
(\Phi_{V\uparrow},\Phi_{V\uparrow})
\ret
&=&\sum_{u,v,r,s}e^{-ik\cdot u}(e^{ik\cdot v}-e^{ik\cdot r})
\kappa_x^{(r)}\varphi_x^{(r)}\kappa_x^{(s)}\varphi_x^{(v)}
(G)_{us}(G^{-1})_{uv}(\Phi_{V\uparrow},\Phi_{V\uparrow})
\ret
&=&L^d \sum_{u,r,s}e^{-ik\cdot u}(1-e^{ik\cdot r})
\kappa_x^{(r)}\varphi_x^{(r)}\kappa_x^{(s)}\varphi_x^{(o)}
(G)_{us}(G^{-1})_{uo}(\Phi_{V\uparrow},\Phi_{V\uparrow}).
\label{nume}
\end{eqnarray}
Summing up (\ref{nume}) over $x$, dividing the result by
(\ref{deno}), and noting that $H_{\rm hop}\Psi_k=0$, we
finally get
\begin{equation}
E_L(k)=\frac{
\sum_{x\in\Lambda}\sum_{u,r,s\in V}e^{-ik\cdot u}(1-e^{ik\cdot r})
U_x\kappa_x^{(r)}\varphi_x^{(r)}\kappa_x^{(s)}\varphi_x^{(o)}
(G)_{us}(G^{-1})_{uo}}
{\sum_{u\in V} e^{-ik\cdot u}(G)_{uo}(G^{-1})_{uo}},
\label{ELk2}
\end{equation}
which converges to an analytic function of $k$ as $L\to\infty$.
By expanding (\ref{ELk2}) in $k$, we find
\begin{eqnarray}
E(k)&=&\frac{
\frac{1}{2}\sum_x\sum_{u,r,s}(k\cdot r)^2
U_x\kappa_x^{(r)}\varphi_x^{(r)}\kappa_x^{(s)}\varphi_x^{(o)}
(G)_{us}(G^{-1})_{uo}}
{\sum_u(G)_{uo}(G^{-1})_{uo}}
+O(|k|^4)
\ret
&=&
J(\lambda)|k|^2+O(|k|^4),
\end{eqnarray}
which is the desired estimate.
\end{proof}

\bigskip\bigskip\noindent
H. T. wishes to thank Tohru Koma for stimulating discussions and
valuable comments, Arisato Kawabata, Kenn Kubo and Koichi
Kusakabe for useful discussions, and Masanori Yamanaka for useful
comments on the manuscript.
A. M. wishes to thank Prof.~Dr.~F.~Wegner and Christian Lantwin
for valuable comments, as well as Herv\'{e} Kunz and
Andr\'as S\"ut\H o for many discussions on the Hubbard model.
We thank Koichi Kusakabe and Hideo Aoki fo\cite{Kusakabe2} prior to
publication.
H. T. also thanks the Meson Science Laboratory at the University of
Tokyo for generously providing access to their computer network.
The research of H. T. is supported in part by Grant-in-Aid for
General Scientific Research from the Ministry of Education,
Science and Culture.

\bigskip
\bigskip
\noindent{\bf Figure Captions}

\bigskip
\noindent
Fig.1
The decorated square lattice.
The hopping matrix elements are given by ${t}_{xy}=t$ for a black
line, ${t}_{xy}=\lambda t$  for a gray line, ${t}_{xx}=4t$ for a site
$x$ of the square lattice, and ${t}_{xx}=\lambda^2t$  for a site $x$
at the middle of a bond, where $t,\lambda>0$.
The on-site Coulomb repulsion is nonvanishing for any site.
It is proved that the ground states exhibit ferromagnetism when the
electron filling factor $\rho$ is not more than and sufficiently close
to ${\rho }_0=1/6$, and paramagnetism when $\rho$ is sufficismall.

\bigskip
\noindent
Fig.2
The defect $D\in\setD[A]$ and nearby clusters.
The white regions represent the occupied sites.
\end{document}